%% file: main.tex
\documentclass[lettersize,journal]{IEEEtran}

\usepackage{amsmath}
\usepackage{amssymb}
\usepackage{amsfonts}
\usepackage{amsthm}
\usepackage{array}
\usepackage{textcomp}
\usepackage{stfloats}
\usepackage{url}
\usepackage{verbatim}
\usepackage{graphicx}
\usepackage{cite}
\usepackage{subfigure}
\usepackage{graphicx}
\usepackage{adjustbox}
\usepackage{booktabs}
\usepackage[skip=5pt]{caption}
\usepackage{multirow}
\usepackage{amssymb}
\usepackage{amsmath}
\usepackage{newtxmath}
\usepackage{diagbox}
\usepackage{url}
\usepackage{xcolor}
\usepackage{xspace}
\usepackage{dsfont}
\usepackage{enumitem}
\usepackage{caption}
\usepackage{balance}
\usepackage{tabularx}
\usepackage{graphicx}

\newtheorem{example}{Example}
\newtheorem{theorem}{Theorem}
\newtheorem{corollary}{Corollary}
\newtheorem{lemma}{Lemma}
\newtheorem{remark}{Remark}
\newtheorem{definition}{Definition}
\newtheorem{prule}{rule}

\newcommand{\myparagraph}[1]{\vspace{1mm} \noindent \textbf{#1}.}
\newcommand{\ie}{{i.e.,}\xspace}

\newcommand{\etal}{et al.\xspace}

\usepackage[ruled, vlined, linesnumbered]{algorithm2e}

\newcommand{\stitle}[1]{\vspace{1ex} \noindent{\bf #1}}

\begin{document}
\title{WCCS: Efficient Wedge Conductance Community Search over Large Temporal Bipartite Graphs (Full Paper)}
	\author{Longlong Lin, Wei Chen, Pingpeng Yuan, Ruikun Luo, Qiangqiang Dai,  Rong-Hua Li
	\IEEEcompsocitemizethanks{
    Longlong Lin and Wei Chen are with the College of Computer and Information Science, Southwest University. Email: longlonglin@swu.edu.cn; WeiChen605@126.com.  Pingpeng Yuan and Ruikun Luo are with Huazhong University of Science and Technology.
		Email: ppyuan@hust.edu.cn; rkluo@hust.edu.cn. Qiangqiang Dai and Rong-Hua Li are with 
		Beijing Institute of Technology.   Email: qiangd66@gmail.com; lironghuabit@126.com}
    }


\maketitle


\begin{abstract}
 Bipartite graphs are ubiquitous for modeling complex interactions between two distinct entity types across numerous practical applications such as e-commerce, academic networks, and social systems. Despite significant progress in community search over bipartite graphs, most prior work is limited to static settings and ignores the rich temporal dynamics present in real-world networks. Moreover, existing methods typically adopt edge-centric measures and strict consecutivity constraints, failing to capture higher-order interactions and frequent yet non-consecutive activities. More importantly, they often neglect the crucial community-quality requirements of both internal cohesiveness and external sparsity, failing to identify critical nodes or including many irrelevant nodes. To address these dilemmas, we propose the novel problem of \emph{Wedge Conductance Community Search (WCCS)}, which aims to identify a query-dependent community that is not only structurally and temporally cohesive but also well-separated from the rest of the network over non-consecutive timestamps. We formalize WCCS by generalizing the classical $(\alpha,\beta)$-core to a higher-order $(\alpha,\beta,\tau)$-wedge core, and by proposing a novel temporal wedge conductance metric that explicitly balances internal density and external sparsity. 
 To solve WCCS efficiently, we first develop an online priority-driven filter-and-expand framework with several effective pruning techniques and a powerful geometric slope optimization for rapid temporal wedge conductance calculation. Subsequently, to further improve scalability, we propose an offline compressed index to accelerate search. Finally, comprehensive experiments on seven real-world datasets demonstrate the effectiveness, efficiency, and scalability of our solutions compared to eight competitors.
\end{abstract}


\bstctlcite{IEEEexample:BSTcontrol}

\input{section/Introduction}

\input{section/Preliminaries}

\input{section/Method1}

\input{section/Method2}

\input{section/Experiment}
\input{section/NewEffectiveness}

\input{section/Relatedwork}
\input{section/Conclusion}

\bibliographystyle{IEEEtran}
\bibliography{main}

\input{section/Appendix}

\clearpage

\end{document}

%% file: section/Introduction.tex
\section{Introduction}
\label{sec:intro}
Bipartite graphs are widely used to model complex relationships between two distinct types of entities, such as customer-product and author-paper networks. With the widespread adoption of bipartite graphs, significant efforts have been dedicated to bipartite graph analysis~\cite{wang2024bipartite}. Among them, community search over bipartite graphs, identifying the high-quality subgraph containing the user-specified query node by simultaneously considering both cohesiveness and bipartite characteristics, which has spurred numerous real-world applications, including recommender systems and fraud detection ~\cite{wangICDE21, wang2023significantbp,wangperbc,Chenpercolation23,LiAWCS22,Xvattrib,Zhangbitruss,ZHOUSize23, zhang2024size, LISize24}. Therefore, numerous community search models have been studied in bipartite graphs based on $(\alpha, \beta)$-core~\cite{wangICDE21, wang2023significantbp}, biclique~\cite{wangperbc, Chenpercolation23}, and bitruss~\cite{Zhangbitruss} in the literature.

\begin{figure}[t] 
	\centering
	\includegraphics[width=1.0\linewidth]{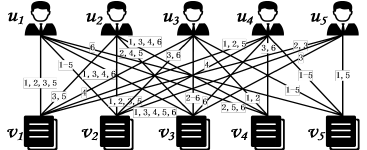}
        \caption{Author-paper temporal bipartite graph. Each edge $(u,v,t)$ indicates that author $u$ published paper $v$ at time $t$.}
	\label{fig:intro}
\end{figure}

Despite their success, most existing models are designed for static networks. Nevertheless, real-world graphs frequently encompass intricate yet valuable temporal information (\textit{e.g.}, in author-paper temporal networks, a temporal edge denotes that an author published a paper at a certain time), which can offer deep insights into the underlying causes and help predict temporal behaviors~\cite{holme2012temporal,cai2023efficient,DBLP:journals/tii/ChenLLJ26,DBLP:conf/icde/LengZQLL25,DBLP:journals/tsmc/LinYLWLJ22,DBLP:journals/tbd/LinYLJ22,DBLP:conf/ppopp/Hou0H00YL00L026,DBLP:conf/asplos/0003WH0DHWY0LYL25,ChenBreahability,li2024historicalabc,wu2024frequency}. One could potentially apply temporal unipartite community search methods~\cite{li2018persistent, wu2014path, huang2015minimum, qin2019mining, qin2020mining, qin2022mining, zhang2023discovering, Ga2018span, aslay2018mining} to bipartite settings by ignoring the bipartite characteristics with two disjoint vertex sets, but such solutions lead to suboptimal empirical results (Section \ref{exp}). For example, they typically measure the subgraph cohesiveness by edge-centric degrees. Nevertheless, the degree based on wedge, a two-hop path in a bipartite graph, outperforms the edge-centric degree~\cite {zhou2021butterfly,cai2023efficient,wang2019vertex,wu2024frequency,li2023persistent,li2024historicalabc}. As shown in Figure \ref{fig:intro}, co-authorship on the same papers captures richer and more meaningful higher-order associations than simply counting the papers published by individual authors. Recently, several studies have explored temporal bipartite community search (Section \ref{sec:rel}). However, they adopted a strict interval constraint, which led to limited flexibility and suboptimal empirical performance (Section \ref{exp}). In contrast, in many real-world systems, many interesting groups appear frequently but irregularly and non-consecutively, such as academic collaboration and fraud rings \cite{shin2017densealert, li2018persistent}. More importantly, their community score only considers the internal cohesiveness and ignores the significant external sparsity,
which is an important factor in measuring the community quality~\cite{DBLP:journals/pacmmod/KamalB24,DBLP:conf/www/LeskovecLM10}.

To this end, we propose the new problem of \emph{\underline{W}edge \underline{C}onductance \underline{C}ommunity \underline{S}earch} (WCCS), which simultaneously captures internal cohesiveness, external sparsity, and temporal frequency. Specifically, WCCS first extends  $(\alpha,\beta)$-core~\cite{liu2020efficient, Luo2023coremain} to higher-order wedge structure, yielding an $(\alpha,\beta)$-wedge core to measure the cohesiveness of subgraphs across snapshots. We further define the temporal wedge conductance (Definition \ref{twc}) to evaluate the community quality, modified from the well-known conductance~\cite{fiedler1973algebraic, DBLP:conf/stoc/SpielmanT04, DBLP:conf/focs/AndersenCL06,DBLP:conf/www/LeskovecLM10,DBLP:conf/aaai/LinLJ23,HeLYLJW24,Linbil2025,LinKDD24,NCSAC2025} (conductance is a traditional measure of community quality, which takes into account both internal and external structural characteristics). As a consequence, given a temporal bipartite graph $\mathcal{G}=(U, V,\mathcal{E})$, a query vertex $q\in U$, a time interval $[t_s,t_e]$, two structural constraints $\alpha, \beta$ and frequency threshold $\tau$, WCCS aims to find a community $C\subseteq U$ (details in Section \ref{sec:pre}) such that $i)$ $C$ is connected and contains $q$, $ii) \ C$ forms an $(\alpha,\beta)$-wedge core with some vertices in $V$ in at least $\tau$ corresponding snapshots within $[t_s,t_e]$, and $iii)$ the temporal wedge conductance $\Phi(C)$ is minimized, which is defined as the maximum average over all sub-segments with length no less than $\tau$ in time-sorted wedge conductance sequence.  The following example further demonstrates our motivations.

\begin{example}\label{example:intro}
Consider the author-paper temporal bipartite graph in Figure \ref{fig:intro}. Suppose we aim to identify a collaborative team centered at $u_3$ within $[1,4]$. Under $\alpha = \beta =2$, the state-of-the-art (SOTA) temporal bipartite community search method \textit{TABC}~\cite{tabc24} evaluates the aggregated graph over the query interval and returns all authors, namely $\{u_1,u_2,u_3,u_4,u_5\}$ which is overly coarse as it only guarantees structural feasibility on the aggregated interval and ignores community quality. The behavior of \textit{PCSearch}~\cite{li2023persistent} is fundamentally different. Under $\tau = 3$, no valid $(2,2)$-persistent core containing $u_3$ exists within $[1,4]$. Hence, PCSearch returns no result, revealing that strict continuity may miss collaborators such as $\{u_1, u_2\}$ that interact with $u_3$ frequently but non-consecutively. In contrast, under our proposed $(\alpha,\beta,\tau)$-wedge core with $\alpha= \beta=2$ and $\tau=3$, $\{u_1,u_2,u_3\}$ is valid at timestamps $t=2, 3, 4$, and thus forms a feasible $(2,2,3)$-wedge core. Further guided by temporal wedge conductance minimization, WCCS returns $\{u_1,u_2,u_3\}$ as a compact community around $u_3$ because it not only satisfies both structural cohesiveness and temporal frequency constraints, but also achieves the minimum temporal wedge conductance value of $0.756$ among all feasible candidate communities.
\end{example}

WCCS is practically important but technically challenging, since it must jointly handle structural cohesiveness, temporal frequency, and community quality. Existing temporal bipartite community models mainly focus on interval-based patterns~\cite{li2023persistent,LiMoreliable,li2024historicalabc}, whereas our model verifies community validity on each snapshot independently. This leads to two main challenges. 
(1) \textit{Challenge 1: Exponential combinatorial search space}. The target result is a cohesive and frequent subgraph containing the query vertex $q$. One needs to enumerate subgraphs containing $q$ across snapshots, verify their frequency constraint, and identify the one with the minimum temporal wedge conductance. Since the number of such subgraphs grows exponentially with both graph size and the number of snapshots, the search space is prohibitively large. 
(2) \textit{Challenge 2: Expensive temporal wedge conductance computation}. During community expansion, temporal wedge conductance must be evaluated repeatedly for candidate communities across multiple snapshots. Moreover, computing this metric requires examining all sorted valid timestamps with a count at least $\tau$, which makes efficient maintenance nontrivial. To address Challenge 1, we develop a frequency-aware filter-and-expand framework with query-driven filtering, priority-guided expansion, effective pruning strategies, and a compressed index for fast candidate retrieval. To address Challenge 2, we transform the computation of temporal wedge conductance into a geometric slope maximization problem on the cumulative wedge conductance curve, and design a dynamic convex-hull maintenance algorithm to avoid redundant recomputation and achieve linear-time updating. The main contributions of the paper are summarized as follows.


\stitle{Novel Model.} To the best of our knowledge, this work is the first to introduce the well-established conductance clustering metric to study the temporal bipartite community search, leading to our novel \emph{\underline{W}edge \underline{C}onductance \underline{C}ommunity \underline{S}earch} (WCCS) model. A key advantage of WCCS is its ability to balance internal and external structural cohesiveness and temporal frequency, whereas previous models cannot.

\stitle{Efficient Online Algorithm.} To efficiently solve the proposed WCCS problem, we develop \underline{P}riority-guided \underline{F}requency-aware \underline{C}ommunity \underline{S}earch (PFCS), an online search algorithm enhanced with effective pruning strategies. Notably, we leverage geometric slope optimization for temporal wedge conductance calculation, which significantly reduces the computational complexity from quadratic to linear.

\stitle{Compressed Offline Index.} To further accelerate query processing, we introduce a compressed offline index structure, the \underline{W}edge \underline{C}ore \underline{I}ndex (WCI). By leveraging this index, all candidate nodes can be retrieved directly, eliminating the need for tedious peeling operations on the raw graph.

\stitle{Comprehensive Experiments.} 
We conduct extensive experiments on seven real-world datasets to systematically evaluate the efficiency and effectiveness of the proposed solutions compared to eight competitors. Our datasets and codes are located at https://anonymous.4open.science/status/WCCS-246E.

%% file: section/Preliminaries.tex
\section{Preliminaries}
\label{sec:pre}

\begin{figure}
	\centering
	\includegraphics[width=\linewidth]{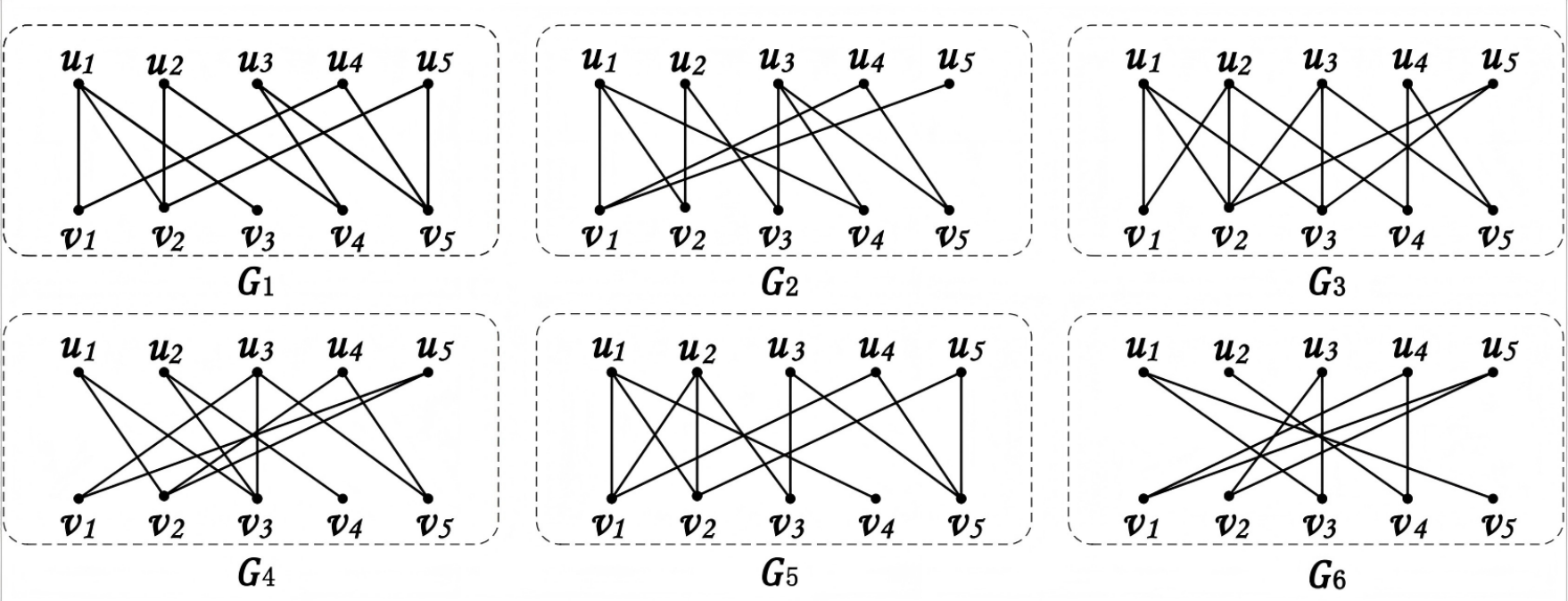}
	\caption{The snapshots of the temporal graph in Figure \ref{fig:intro}.}
	\label{fig:snapshot}
\end{figure}

\subsection{Basic Notations} 
Let $\mathcal{G}=(U, V,\mathcal{E})$ be an undirected temporal bipartite graph, in which $U$ and $V$ are two disjoint vertex sets (\ie $U \cap V = \emptyset$). Each temporal edge $(u,v,t) \in \mathcal{E}$ is an interaction between $u \in U$ and $v \in V$ at timestamp $t$. Two temporal edges $(u,v,t_i)$ and $(u,v,t_j)$ are treated as distinct if $t_i \neq t_j$, meaning $u$ and $v$ interact at multiple different timestamps. Let $\mathcal{T}=\{t|(u,v,t)\in \mathcal{E}\}$ be the set of all timestamps. 
For any timestamp $t\in \mathcal{T}$, the snapshot of $\mathcal{G}$ at  $t$ is $G_t =(U, V, E_t)$, where $E_t =\{(u,v)| (u,v,t) \in \mathcal{E}\}$. For any vertex $u \in U$ and $t\in \mathcal{T}$, its incident neighbor set is $N_t(u) = \{v \in V \mid (u, v) \in E_t\}$ and its degree is denoted $d_t(u) = |N_t(u)|$. Similarly, we can define the incident neighbors $N_t(v)$ of $v\in V$. To capture the higher-order structure in bipartite graphs, we denote the \textbf{wedge} (a wedge is a path of length 2) \textbf{neighbors} of $u$ in $G_t$ as $W_t(u)  = \{u' \in U \setminus \{u\} \mid \exists v \in V, (u, v) \in E_t \wedge (u', v) \in E_t\}$. We further define the \textbf{wedge degree} of $u$ in $G_t$ as $d_w(u,G_t)=\sum_{v\in N_t(u)}(d_t(v)-1)$,
which counts the number of wedges incident to $u$ at timestamp $t$. 

Following \cite{wu2024frequency}, in this paper, we also aim to identify a unilateral\footnote{co-clustering \cite{DBLP:journals/tcyb/XueNLWL24}, which simultaneously clusters both types of nodes, is another interesting yet orthogonal problem that we leave for future work.} community containing the query node, where a unilateral community refers to a subset of nodes of the same type \cite{wu2024frequency}. Without loss of generality, we assume the query vertex $q$ belongs to $U$. Therefore, we introduce the  $(\alpha,\beta)$-wedge core with the same type, extending the well-established \((\alpha,\beta)\)-core \cite{alphabetacoreCIKM,liu2020efficient} from edges to wedges to better capture the unique higher-order features of bipartite graphs.

\begin{definition}[$(\alpha,\beta)$-Wedge Core]\label{wc}
Given a snapshot $G_t=(U, V, E_t)$ and two positive integers $\alpha$ and $\beta$, a vertex subset $C \subseteq U$ is an $(\alpha,\beta)$-wedge core if there exists a  subset $V_C \subseteq V$ such that the induced subgraph $S=(C, V_C, E_S)$ of $G_t$ satisfies: (i) $\forall u \in C$, $u$ has at least distinct $\alpha$ wedge neighbors in $C$ through vertices in $V_C$, i.e., $|\{u' \in C \setminus \{u\} \mid \exists v \in V_C, (u, v) \in E_t \wedge (u', v) \in E_t\}|\geq \alpha$; (ii)  $\forall v \in V_C$, $v$ has at least distinct $\beta$ wedge neighbors in $V_C$ through vertices in $C$. Namely, $|\{v' \in V_C \setminus \{v\} \mid \exists u \in C, (u, v) \in E_t \wedge (u, v') \in E_t\}|\geq \beta$; (iii) $S$ is connected. 
\end{definition}

\begin{example}
Figure \ref{fig:snapshot} shows all snapshots of Figure \ref{fig:intro}. In snapshot $G_3$, the wedge neighbors of $u_1$ are $\{u_2,u_3,u_5\}$, and its wedge degree is $d_w(u_1,G_3)=1+3+2=6$. Consider $C=\{u_1,u_2,u_3\}$. There exists a support set $V_C=\{v_1,v_2,v_3,v_4,v_5\}$ such that each $u\in C$ has at least $2$ wedge neighbors in $C$, each $v\in V_C$ has at least $2$ wedge neighbors in $V_C$, and the induced subgraph $S=(C,V_C,E_S)$ is connected. Hence, $C$ is a $(2,2)$-wedge core in $G_3$.
\end{example}



\subsection{Problem Formulation}

\begin{definition}[Candidate Timestamp]\label{CT}
 Given a temporal bipartite graph $\mathcal{G} = (U, V, \mathcal{E})$, for any vertex $u \in U$, we define $\mathcal{CT}(u) \subseteq \mathcal{T}$ as the set of candidate timestamps where $u$ is contained in the maximal $(\alpha, \beta)$-wedge core in $G_t$. Consider a vertex set $C \subseteq U$, the candidate timestamp set of $C$ is defined as $\mathcal{CT}_{C} = \bigcap_{u \in C} \mathcal{CT}(u).$
\end{definition}

\begin{definition}[Valid Timestamp]\label{vt}
Given a temporal bipartite graph $\mathcal{G} = (U, V, \mathcal{E})$ and a vertex set $C \subseteq U$, a timestamp $t \in \mathcal{T}$ is a valid timestamp if $C$ is an $(\alpha, \beta)$-wedge core in $G_t$. The set of all valid timestamps of $C$ is denoted by $\mathcal{T}_{C}$. Clearly, $\mathcal{T}_{C} \subseteq \mathcal{CT}_{C}$ for any $C \subseteq U$.
\end{definition}

\begin{definition}[$(\alpha, \beta, \tau)$-Wedge Core]\label{abc}
Given a temporal bipartite graph $\mathcal{G} = (U, V, \mathcal{E})$ and  frequency  threshold $\tau$, a vertex set $C \subseteq U$ is called an $(\alpha, \beta, \tau)$-wedge core if $C$ forms an $(\alpha, \beta)$-wedge core in at least $\tau$ valid timestamps, i.e., $|\mathcal{T}_{C}| \geq \tau$.
\end{definition}

To measure the quality of communities in a bipartite graph, we extend the well-known conductance \cite{fiedler1973algebraic, DBLP:conf/stoc/SpielmanT04, DBLP:conf/focs/AndersenCL06,DBLP:conf/www/LeskovecLM10,DBLP:conf/aaai/LinLJ23} concept to bipartite settings via wedges.

\begin{definition}[Wedge Conductance]
\label{def:wc}
Given a snapshot $G_t=(U,V,E_t)$ and a node set $C\subseteq U$, let $\bar{C}=U\setminus C$. The wedge conductance of $C$ at timestamp $t$ is defined as
\begin{equation}
\phi_w(C,t)=\frac{\mathrm{cut}_w(C,t)}
{\min\!\bigl(\mathrm{vol}_w(C,t),\,\mathrm{vol}_w(\bar{C},t)\bigr)},
\end{equation}
\begin{equation}
\mathrm{cut}_w(C,t)=\sum_{u\in C}\sum_{v\in N_t(u)} |N_t(v)\cap \bar{C}|,
\end{equation}
\begin{equation}
\mathrm{vol}_w(C,t)=\sum_{u\in C} d_w(u,G_t).
\end{equation}
\end{definition}
Intuitively, $\mathrm{cut}_w(C,t)$ counts the wedges crossing the boundary between $C$ and $\bar{C}$, while $\mathrm{vol}_w(C,t)$ measures the total wedge volume incident to $C$. We further incorporate temporal information to define a frequency-aware temporal wedge conductance for a community over its valid timestamps.

\begin{definition}[Temporal Wedge Conductance]\label{twc}
Given a temporal bipartite graph $\mathcal{G} = (U, V, \mathcal{E})$, an $(\alpha, \beta, \tau)$-wedge core $C \subseteq U$, and its sorted valid timestamps $\mathcal{T}_C = \{t_1, t_2, ..., t_m\}$, the temporal wedge conductance of $C$ is defined as
\begin{equation}
\Phi(C) = \max_{\substack{1 \le i \le j \le m \\ \ j - i + 1 \ge \tau}} \frac{1}{j - i + 1} \sum_{k=i}^{j} \phi_w(C, t_k).
\end{equation}
\end{definition}

\noindent \textbf{Problem Definition.} Given a temporal bipartite graph $\mathcal{G} = (U, V, \mathcal{E})$, a query node $q \in U$, a query time interval $\mathcal{T}_Q = [t_s, t_e]$, structural constraints $\alpha$, $\beta$, and a frequency threshold $\tau$, the \textbf{W}edge \textbf{C}onductance \textbf{C}ommunity \textbf{S}earch (WCCS) aims to find a vertex set $C \subseteq U$ such that: (i) $q \in C$; (ii) $C$ is an $(\alpha, \beta, \tau)$-wedge core within $\mathcal{T}_Q$; and (iii) the temporal wedge conductance $\Phi(C)$ is minimized.

The objective is designed to capture robust temporal quality rather than a single global average over all valid timestamps. Specifically, we minimize the maximum average wedge conductance over all contiguous subsegments of the sorted valid timestamp sequence with length at least $\tau$. Such a max-average formulation ensures that the returned community remains well separated even in its weakest sustained period, which aligns with the worst-case principle widely adopted in robust optimization~\cite{Beyerrobust07, Gor15}. In contrast, a $\min$-based objective would only preserve the best-performing period and could entirely overlook long weak intervals, while a global average over all valid timestamps may mask substantial local deterioration by averaging it with more favorable timestamps.

\begin{example}
Consider the temporal bipartite graph shown in Figure \ref{fig:intro} with query interval $[1,6]$. Suppose the query node is $u_1$, $\alpha=\beta=2$, and $\tau=3$. The vertex set $C=\{u_1,u_2,u_3\}$ is a $(2,2)$-wedge core at timestamps $\{2,3,4\}$ according to Figure \ref{fig:snapshot}. In $G_2$, the wedge volumes are $\mathrm{vol}_w(C,2)=9$ and $\mathrm{vol}_w(\bar{C},2)=5$, and the wedge cut is $\mathrm{cut}_w(C,2)=3$. Hence, $\phi_w(C,2)=\frac{3}{\min(9,5)}=\frac{3}{5}=0.600.$ Similarly, $\phi_w(C,3)=1.000$ on $G_3$, and $\phi_w(C,4)=\frac{2}{3}\approx 0.667$ on $G_4$. Therefore, the conductance sequence over the valid timestamps is $\{\phi_w(C,2)=0.600,\ \phi_w(C,3)=1.000,\ \phi_w(C,4)=0.667\}.$ With $\tau=3$, only the entire valid timestamp set $\{2,3,4\}$ satisfies the length constraint, and its temporal wedge conductance is $\frac{0.600+1.000+0.667}{3}=0.756.$ Finally, after enumerating all feasible communities containing $u_1$, the optimal community $C^*=\{u_1,u_4,u_5\}$ is returned, with valid timestamps $\{1,2,5\}$ and $\Phi(C^*)=0.672$.
\end{example}

\begin{theorem}
\label{th:NP-Hard}
The proposed WCCS problem is NP-hard. Furthermore, unless $P=NP$, there exists no polynomial-time approximation algorithm for the problem.
\end{theorem}

%% file: section/Method1.tex
\section{Online Search Algorithm for WCCS}
\label{sec:method1}
This section presents an efficient online search algorithm, PFCS (\underline{P}riority-guided \underline{F}requency-aware \underline{C}ommunity \underline{S}earch), which incrementally explores the community structure w.r.t both topological cohesiveness and temporal frequency.

\subsection{Overview}
\label{subsec:framework}
Our online algorithm, $PFCS$, operates under a progressive filter-and-expand paradigm. The core idea is to anchor the search at query node $q$ and iteratively expand candidate communities that satisfy the $(\alpha, \beta, \tau)$-wedge core condition across multiple time snapshots. Each intermediate community is dynamically evaluated using the proposed temporal wedge conductance, while unpromising expansions are pruned by time-aware and quality-aware strategies. Specifically, $PFCS$ begins with \underline{\textbf{Candidate Set Filtering}}, which discards vertices that can never be contained in final result, thereby significantly reducing the search space before expansion. To accelerate repeated quality evaluation, an incremental \underline{\textbf{Temporal Wedge Conductance Calculation}} for any candidate community $C$ is proposed to transform the complex metric evaluation into a geometric slope maximization problem on the cumulative wedge conductance curve. This enables constant-amortized maintenance of $\Phi(C)$ without exhaustive enumeration and calculation. Based on the above strategies, \underline{\textbf{Candidate Community Expansion}} organizes the search as a best-first exploration rooted at $C_0=\{q\}$, where each state corresponds to a candidate community together with its valid timestamps $\mathcal{T}_C\subseteq \mathcal{T}_Q$ and Temporal Wedge Conductance (Section \ref{sec:pre}). Guided by the computed priority, $PFCS$ iteratively expands the current community $C$ by including the most promising neighbor $u'$, while strictly verifying the $(\alpha, \beta)$-wedge core property and applying aggressive pruning to prune unpromising branches.

\subsection{Candidate Set Filtering}
\label{subsec:fliter}

Given the query interval $\mathcal{T}_Q=[t_s,t_e]$, for any timestamp $t\in \mathcal{T}_Q$, let $\mathcal{W}_{\alpha,\beta}(G_t)$ denote the maximal $(\alpha,\beta)$-wedge core in $G_t$. We use $\mathcal{W}_{\alpha,\beta}(G_t,q)$ to denote the connected component of $\mathcal{W}_{\alpha,\beta}(G_t)$ that contains the query node $q$. If $q\notin \mathcal{W}_{\alpha,\beta}(G_t)$, then $\mathcal{W}_{\alpha,\beta}(G_t,q)=\emptyset$. 

\begin{lemma}[Query-anchored filtering rule]
\label{le:qa-filter}
Given a temporal bipartite graph $\mathcal{G}=(U,V,\mathcal{E})$, a query node $q\in U$, and parameters $\alpha$, $\beta$, and $\tau$, if a community $C\subseteq U$ is a valid $(\alpha,\beta,\tau)$-wedge core containing $q$, then  $\forall u\in C$ must appear together with $q$ in the same $(\alpha,\beta)$-wedge core component in at least $\tau$ snapshots within $\mathcal{T}_Q$, i.e.,$
\bigl|\{t\in \mathcal{T}_Q \mid u\in \mathcal{W}_{\alpha,\beta}(G_t,q)\}\bigr|\ge \tau.$
\end{lemma}

Guided by Lemma \ref{le:qa-filter}, Algorithm \ref{alg:abocore} derives the query-anchored $(\alpha,\beta,\tau)$-wedge core candidate set. It first restricts $\mathcal{G}$ to the query interval and initializes two frequency counters, $cntU[\cdot]$ and $cntV[\cdot]$, for vertices in $U$ and $V$, respectively (Lines 1--3). For each timestamp $t\in[t_s,t_e]$, \textit{PeelWC} is invoked to extract $\mathcal{W}_{\alpha,\beta}(G_t,q)$, i.e., the connected component containing $q$ in the maximal $(\alpha,\beta)$-wedge core of $G_t$. Then, for every $w\in \mathcal{W}_{\alpha,\beta}(G_t,q)$, its counter is increased by one (Lines 4--7). After all snapshots have been scanned, the query-side candidate set is obtained as $U^{cand}=\{u\in U\mid cntU[u]\ge \tau\}$, while the support-side candidate set is obtained as $V^{cand}=\{v\in V\mid cntV[v]\ge 1\}$. The final candidate subgraph is induced by $U^{cand}\cup V^{cand}$ (Lines 8--11). Internally, \textit{PeelWC} repeatedly removes vertices violating the wedge core constraints until convergence. If $q$ survives, it returns the connected component containing $q$. Otherwise, it returns $\emptyset$ (Lines 12--20). Therefore, the total cost of Algorithm \ref{alg:abocore} is the sum of the snapshot-wise peeling costs over the query interval.

\begin{algorithm}[t]
\caption{\textsf{WCore} $(\mathcal{G}, q, [t_s, t_e], \alpha, \beta, \tau)$}
\label{alg:abocore}
\footnotesize
\SetKwInOut{Input}{Input}
\SetKwInOut{Output}{Output}
\SetAlgoVlined
\SetKwFunction{PeelWC}{PeelWC}
\SetKwFunction{BFS}{BFS}
\Input{$\mathcal{G}=(U,V,\mathcal{E})$: temporal bipartite graph; $q$: query node;\\ $\alpha,\beta,\tau$: structural and frequency constraints; $[t_s, t_e]$: query interval}
\Output{Candidate subgraph $\mathcal{G}_{cand}$}

$\mathcal{G}_{[t_s,t_e]} \leftarrow \{(U,V,E_t) \mid t \in [t_s,t_e]\}$\;
$cntU[u] \leftarrow 0,\ \forall u \in U$\;
$cntV[v] \leftarrow 0,\ \forall v \in V$\;
\For{$t\in[t_s,t_e]$}{
    $\mathcal{W}_{\alpha,\beta}(G_t, q) \leftarrow \PeelWC(G_t, q, \alpha, \beta)$\;
    $cntU[u] \leftarrow cntU[u] + 1,\ \forall u \in \mathcal{W}_{\alpha,\beta}(G_t, q)\cap U$\;
    $cntV[v] \leftarrow cntV[v] + 1,\ \forall v \in \mathcal{W}_{\alpha,\beta}(G_t, q)\cap V$\;
}
$U^{cand} \leftarrow \{u \in U \mid cntU[u] \ge \tau\}$\;
$V^{cand} \leftarrow \{v \in V \mid cntV[v] \ge 1\}$\;
$\mathcal{G}_{cand} \leftarrow \mathcal{G}_{[t_s,t_e]}[U^{cand}\cup V^{cand}]$\;
\Return{$\mathcal{G}_{cand}$}\;

\vspace{0.2cm}
\textbf{Function} \PeelWC{$(G_t,q,\alpha,\beta)$}:\\
$H \leftarrow G_t$\;
$S \leftarrow \{u \in U(H) \mid |W_t(u,H)| < \alpha\} \cup \{v \in V(H) \mid |W_t(v,H)| < \beta\}$\;
\While{$S \neq \emptyset$}{
    $H \leftarrow H \setminus S$\;
    \lIf{$q \notin U(H)$}{\Return{$\emptyset$}}
    $S \leftarrow \{u \in U(H) \mid |W_t(u,H)| < \alpha\} \cup \{v \in V(H) \mid |W_t(v,H)| < \beta\}$\;
}
$\mathcal{W}_{\alpha,\beta}(G_t, q) \leftarrow \BFS(H, q)$\;
\Return{$\mathcal{W}_{\alpha,\beta}(G_t, q)$}\;
\end{algorithm}

\begin{example}
Reconsider the six snapshots in Figure \ref{fig:snapshot}, where we perform community search over $[1,6]$. Let $\alpha=3$, $\beta=2$, $\tau=4$, and let the query node be $u_3$. For each $t\in\{1,\ldots,6\}$, we invoke \textsc{PeelWC} to obtain the query-anchored $(3,2)$-wedge core component containing $u_3$ in $G_t$, denoted by $\mathcal{W}_{3,2}(G_t,u_3)$. For example, $u_2$ belongs to the same query-anchored component as $u_3$ at timestamps $t=1,3,5$, and thus $cntU[u_2]=3$. Since $cntU[u_2]<\tau=4$, $u_2$ is filtered out. In contrast, $u_1$ appears in the query-anchored component at timestamps $t=1,3,4,5$, yielding $cntU[u_1]=4$. After all six snapshots are processed, applying the frequency threshold $\tau=4$ on the query side yields the candidate set $U^{cand}=\{u_1,u_3,u_5\}$, which is then expanded in the subsequent stage.
\end{example}

\subsection{Temporal Wedge Conductance Calculation}
\label{sub:Cal}

According to Definition \ref{twc}, the temporal wedge conductance $\Phi(C)$ is the maximum average of $\phi_w(C,t)$ over all contiguous segments in the sorted valid timestamp sequence whose lengths are at least $\tau$. Let $\mathcal{T}_C=\{t_1,\dots,t_L\}$ be the sorted valid timestamps of $C$, and let the wedge conductance sequence of $C$ be denoted by $WCS[C]=\bigl[\phi_w(C,t_1),\phi_w(C,t_2),\ldots,\phi_w(C,t_L)\bigr].$ Here, $WCS[C][k]$ denotes the $k$-th element of the sequence $WCS[C]$. A naive computation needs to inspect all contiguous subsequences $(t_i,\ldots,t_j)$ satisfying $j-i+1\ge\tau$, which incurs a time complexity of $O(L^2)$. Since this computation is repeatedly invoked during community expansion, we transform it into a geometric slope maximization problem and reduce the cost to linear time.

\begin{definition}[Cumulative Wedge Conductance Curve]
Given the wedge conductance sequence $WCS[C]$ of length $L$, the cumulative wedge conductance curve of $C$, abbreviated as $CWC[C]$, is defined as $CWC[C][\ell]=\sum_{k=1}^{\ell}WCS[C][k]$ for $1\le \ell\le L$
with $CWC[C][0]=0$. Here, $CWC[C][\ell]$ denotes the $\ell$-th element of the sequence $CWC[C]$.
\end{definition}

\begin{figure}[t]
\centering
\includegraphics[width=1.0\linewidth]{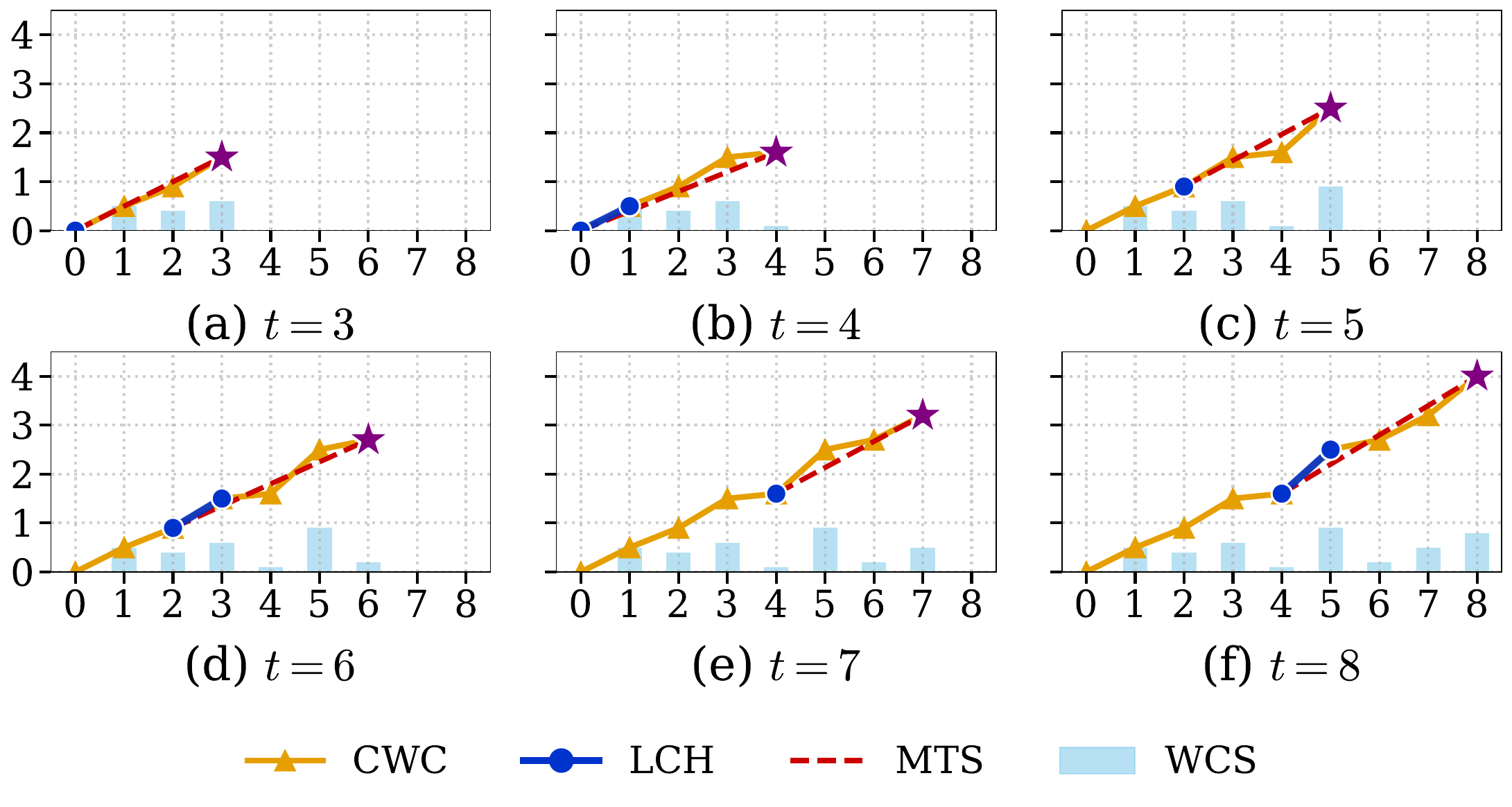}
\caption{Running example of computing temporal wedge conductance for a wedge conductance sequence $[0.5,0.4,0.6,0.1,0.9,0.2,0.5,0.8]$ with $\tau=3$. The x-axis and y-axis represent the prefix index and the cumulative value $CWC$, respectively.}
\label{fig:MTS}
\end{figure}

The sequence $CWC[C]$ can be embedded into a two-dimensional plane by mapping each prefix index $\ell$ to the point $(\ell,CWC[C][\ell])$, as illustrated in Figure \ref{fig:MTS}. For any contiguous subsequence $(t_i,\ldots,t_j)$ with $1\le i\le j\le L$, the average wedge conductance equals the geometric slope between the points $(i-1,CWC[C][i-1])$ and $(j,CWC[C][j])$, i.e.,
$
\frac{1}{j-i+1}\sum_{k=i}^{j}WCS[C][k]
=
\frac{CWC[C][j]-CWC[C][i-1]}{j-i+1}
=
\mathrm{slope}_C(i-1,j).
$ Thus, computing $\Phi(C)$ is equivalent to finding, for each endpoint $j$, the feasible starting index that maximizes the slope.

\begin{definition}[Maximum $\tau$-truncated Slope]
\label{de:mts}
Given the curve $CWC[C]$ with valid-time length $L\ge \tau$, the maximum $\tau$-truncated slope for an endpoint $j\in[\tau,L]$ is defined as
\[
MTS[C][j]=\max\{\mathrm{slope}_C(k,j)\mid 0\le k\le j-\tau\}.
\]
\end{definition}

Intuitively, $MTS[C][j]$ is the maximum average wedge conductance over all sorted valid time windows ending at $j$ with length at least $\tau$. For any community $C$ with at least $\tau$ valid timestamps, its temporal wedge conductance can be directly expressed as $\Phi(C)=\max\{MTS[C][j]\mid j\in[\tau,L]\}.$ Thus, the final value of $\Phi(C)$ can be obtained by taking the maximum over all feasible values of $MTS[C][j]$. The main challenge is to maintain $MTS[C][j]$ as $j$ increases, which requires efficiently finding the optimal starting index in $[0,j-\tau]$. To this end, we employ a convex-hull-based optimization strategy.

\begin{definition}[Valid Lower Convex Hull]
\label{LCH}
For an endpoint $j$, the feasible starting indices range from $0$ to $j-\tau$. The valid lower convex hull, denoted by $LCH_j(C)$, is defined as a subsequence of indices $\{p_1,p_2,\dots\}\subseteq[0,j-\tau]$ such that the slopes between consecutive points strictly increase, i.e., $
\mathrm{slope}_C(p_m,p_{m+1})<\mathrm{slope}_C(p_{m+1},p_{m+2}).$
\end{definition}

The hull $LCH_j(C)$ contains all candidate starting indices that may yield the maximum slope to endpoint $j$. Geometrically, any starting point not lying on $LCH_j(C)$ is dominated by hull points and cannot produce a larger slope than the best hull point. When the endpoint increases from $j$ to $j+1$, the value of $MTS[C][j+1]$ can be updated from the current hull together with the newly appended point $(j+1,CWC[C][j+1])$, as stated in the following lemmas.

\begin{lemma}
\label{le:convex}
Let $a,b$ be the last two indices in the current hull, where $a<b$. When a new point $(c,CWC[C][c])$ is appended, the point $(b,CWC[C][b])$ violates convexity and must be removed if $\mathrm{slope}_C(b,c)\le \mathrm{slope}_C(a,b)$.
\end{lemma}

\begin{lemma}
\label{le:head}
For a fixed endpoint $j$, if $\mathrm{slope}_C(a,j)\le \mathrm{slope}_C(b,j)$, then index $a$ can never outperform $b$ for any later endpoint, and thus $(a,CWC[C][a])$ should be removed from the head of the hull.
\end{lemma}

Based on Lemma \ref{le:convex} and Lemma \ref{le:head}, we design Algorithm \ref{alg:MTS} to dynamically maintain the lower convex hull and compute $MTS[C][j]$ for each endpoint $j$. Specifically, it first checks whether $C$ satisfies the frequency constraint and then constructs the cumulative wedge conductance curve $CWC[C]$ from the sequence $WCS[C]$ (Lines 1--6). Next, for each endpoint $j$ from $\tau$ to $L$, it incrementally updates the hull according to the above two lemmas, retrieves the current optimal starting index from the head, computes $MTS[C][j]$, and updates $\Phi(C)$ accordingly (Lines 8--18). The hull is maintained as a monotone deque with head and tail pointers $i_s$ and $i_e$ according to Lemma \ref{le:convex} and Lemma \ref{le:head} (Lines 10--15). Since each index enters and leaves the hull at most once, the overall time complexity is $O(L)$.

\begin{algorithm}[t]
\footnotesize
\SetKwInOut{Input}{Input}
\SetKwInOut{Output}{Output}
\SetAlgoVlined
\caption{\textsf{ComputeMTS}$(C, WCS[C], \tau)$}
\label{alg:MTS}
\Input{$C$: candidate community; $\tau$: frequency constraint; $WCS[C]$: wedge conductance sequence of $C$}
\Output{Temporal wedge conductance $\Phi(C)$}

$L \leftarrow |WCS[C]|$\;
\If{$L < \tau$}{\Return{$+\infty$}\;}
$CWC[C][0] \leftarrow 0$\;
\For{$\ell \leftarrow 1$ \KwTo $L$}{
    $CWC[C][\ell] \leftarrow CWC[C][\ell-1] + WCS[C][\ell]$\;
}
$i_s \leftarrow 0,\ i_e \leftarrow -1,\ LCH \leftarrow \emptyset,\ \Phi(C) \leftarrow -\infty$\;

\For{$j \leftarrow \tau$ \KwTo $L$}{
    $s \leftarrow j-\tau$\;
    \While{$i_s < i_e \land \mathrm{slope}_C(LCH[i_e],s) \le \mathrm{slope}_C(LCH[i_e-1],LCH[i_e])$}{
        $i_e \leftarrow i_e - 1$\;
    }
    $i_e \leftarrow i_e + 1$;\ $LCH[i_e] \leftarrow s$\;
    \While{$i_s < i_e \land \mathrm{slope}_C(LCH[i_s],j) \le \mathrm{slope}_C(LCH[i_s+1],j)$}{
        $i_s \leftarrow i_s + 1$\;
    }
    $i^* \leftarrow LCH[i_s]$\;
    $MTS[C][j] \leftarrow \mathrm{slope}_C(i^*,j)$\;
    \If{$MTS[C][j] > \Phi(C)$}{
        $\Phi(C) \leftarrow MTS[C][j]$\;
    }
}
\Return{$\Phi(C)$}\;
\end{algorithm}

\begin{example}
Figure \ref{fig:MTS} illustrates the computation of the maximum $\tau$-truncated slope for a candidate community $C$ with $WCS[C]=[0.5,0.4,0.6,0.1,0.9,0.2,0.5,0.8]$ and $\tau=3$. The corresponding cumulative sequence is $CWC[C]=[0,0.5,0.9,1.5,1.6,2.5,2.7,3.2,4.0]$. The process starts from $j=3$, since shorter windows do not satisfy the frequency constraint. When $j=5$, index $1$ is removed by Lemma \ref{le:convex} because $\mathrm{slope}_C(1,2)=0.4<\mathrm{slope}_C(0,1)=0.5$, and index $0$ is further removed from the head by Lemma \ref{le:head} since $\mathrm{slope}_C(0,5)=0.5<\mathrm{slope}_C(2,5)\approx0.53$. The same maintenance continues for later endpoints. Finally, the algorithm returns $\Phi(C)=\max_j MTS[C][j]=0.6$, corresponding to the segment from index $4$ to $8$.
\end{example}

\subsection{Candidate Community Expansion}
\label{subsec:expansion}
Following the query-anchored filtering, the search space is reduced to $\mathcal{G}_{cand}$. Since WCCS requires examining all $(\alpha,\beta)$-wedge cores containing $q$, a natural approach is to enumerate candidate communities from $q$ on $\mathcal{G}_{cand}$, following the spirit of the classical Bron{-}Kerbosch framework \cite{bron1973algorithm}. However, directly exploring all feasible supersets still suffers from severe combinatorial explosion. Therefore, we organize the expansion as a best-first search and use a conductance-aware heuristic to prioritize promising vertices. This allows high-quality communities to be identified earlier, thereby tightening the global bound and improving pruning effectiveness.

\begin{figure*}[t]
\centering
\includegraphics[width=0.98\textwidth]{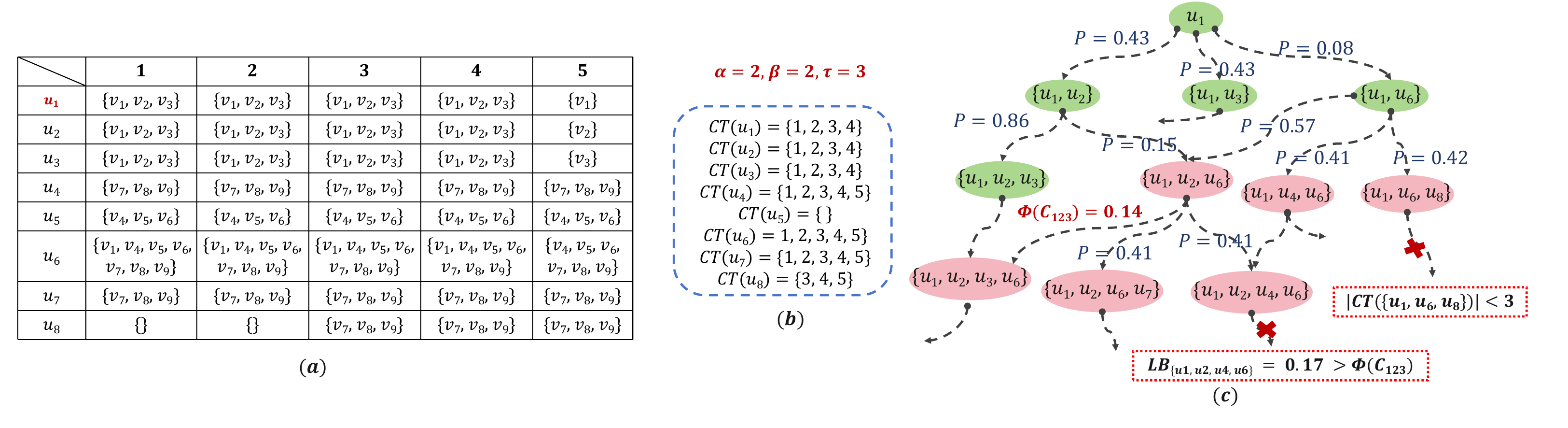}
\caption{Running example of best-first expansion and pruning in WCCS with $\alpha=2$, $\beta=2$, $\tau=3$, and query $q=u_1$. The structure of the temporal bipartite graph is shown in $(a)$, and $(b)$ shows the candidate timestamp sets of the vertices. The search process is detailed in $(c)$, where each search-tree node expands by adding wedge neighbors and is annotated with $\mathcal{CT}$, candidate priorities, and the key values used for pruning. The community $\{u_1,u_2,u_3\}$ achieves the optimal quality $0.14$, whereas other branches demonstrate CT-frequency pruning, wedge core frequency pruning, early-stop in conductance evaluation, and branch lower-bound pruning.}
\label{fig:expansion}
\end{figure*}

\begin{definition}[Conductance-based Vertex Priority]
\label{def:priority}
Given a snapshot $G_t=(U,V,E_t)$, a current community $C\subseteq U^{cand}$, and a candidate vertex $x\in U^{cand}\setminus C$, let $\mathrm{cut}_w(x,C,t)$ denote the number of wedges incident to $x$ whose other endpoint lies in $C$, and let $d_w(x,G_t)$ denote the wedge degree of $x$ in $G_t$. The snapshot priority of $x$ with respect to $C$ at timestamp $t$ is defined as $\mathrm{P}_t(x\mid C)=\frac{\mathrm{cut}_w(x,C,t)}{\max\{1,d_w(x,G_t)\}}.$ Since the exact valid timestamps of $C\cup\{x\}$ are still unknown, we aggregate the snapshot priorities over the candidate timestamp set $\mathcal{CT}_{C\cup\{x\}}$ to evaluate the overall contribution of $x$ across timestamps. Accordingly, the aggregate priority of $x$ with respect to $C$ is defined as
\begin{equation}
 \mathcal{P}(x\mid C)=
\frac{\sum\limits_{t\in \mathcal{CT}_{C\cup \{x\}}} \mathrm{P}_t(x\mid C)}
{|\mathcal{CT}_{C\cup \{x\}}|}.   
\end{equation}
\end{definition}
The snapshot priority $\mathrm{P}_t(x\mid C)$ measures the local cohesiveness gain at timestamp $t$ after inserting $x$ into $C$. By averaging over $\mathcal{CT}_{C\cup\{x\}}$, $\mathcal{P}(x\mid C)$ favors candidates that are likely to maintain stable and high-quality performance across snapshots. Guided by this metric, the algorithm explores candidate vertices in descending order of $\mathcal{P}(x\mid C)$ during expansion. Once $x$ is selected and the community is expanded to $C'=C\cup\{x\}$, the snapshot priorities of the remaining candidates are updated incrementally rather than being recomputed from scratch. Specifically, for any $y\in U^{cand}\setminus C'$, the number of shared wedges updates as $\mathrm{cut}_w(y,C',t)=\mathrm{cut}_w(y,C,t)+|N_t(y)\cap N_t(x)|$, while $d_w(y,G_t)$ remains unchanged.

\stitle{CheckWCore Procedure.}
A crucial and frequent operation during expansion is verifying whether the augmented community satisfies the $(\alpha,\beta)$-wedge core constraints. A naive approach would perform a global peeling to recompute the support vertex set at every expansion step, which is prohibitively expensive. Instead, we devise an incremental local verification strategy based on the monotonicity of support sets.

\begin{lemma}
\label{mono-support}
Let $V_t(C)\subseteq V^{cand}$ denote the maximal support set at timestamp $t$ such that each $v\in V_t(C)$ has at least $\beta$ wedge neighbors with respect to $C$ in $G_t$. When $C$ expands to $C'=C\cup\{x\}$, it holds that $V_t(C)\subseteq V_t(C')$. Consequently, the wedge degree of every vertex $u\in C$ is non-decreasing.
\end{lemma}

Guided by Lemma \ref{mono-support}, the feasibility of $C'$ can be verified locally at each timestamp. Since all vertices in $C$ that already satisfy the $\alpha$-constraint remain valid after the expansion, it is unnecessary to re-check them. Based on this observation, the \textit{CheckWCore} procedure first updates the validity of support vertices adjacent to $x$ in $G_t$, and then verifies whether the newly added vertex $x$ satisfies the $\alpha$-constraint with respect to the updated support set $V_t(C')$. Consequently, the verification cost is reduced from recomputing a global support set to a local update around $x$.

\myparagraph{Incremental Wedge Conductance Update}
After structural validation via \textit{CheckWCore}, the next step is to evaluate the quality of the expanded community $C'$. According to Definition \ref{def:wc}, the main cost comes from repeatedly evaluating $\mathrm{vol}_w(C',t)$ and $\mathrm{cut}_w(C',t)$ for each expansion state. To avoid redundant traversals, we analyze the topological changes caused by inserting a new vertex. For any $x\in U^{cand}\setminus C$ inserted into $C$, the wedge volume updates as
\begin{equation}
\mathrm{vol}_w(C',t)=\mathrm{vol}_w(C,t)+d_w(x,G_t).
\label{eq:vol-update}
\end{equation}

\begin{lemma}
\label{eq:cut-update}
Given a snapshot $G_t$, the wedge cut of the expanded community $C'=C\cup\{x\}$ satisfies
\begin{equation}
\mathrm{cut}_w(C',t)=\mathrm{cut}_w(C,t)+d_w(x,G_t)-2\,\mathrm{cut}_w(x,C,t).
\end{equation}
\end{lemma}

\begin{corollary}
\label{eq:con-update}
Given a candidate community $C$ and any vertex $x\in U^{cand}\setminus C$ in $G_t$, the wedge conductance of the expanded community $C'=C\cup\{x\}$ is
\begin{equation}
\phi_w(C',t)=
\frac{
\mathrm{cut}_w(C,t)+d_w(x,G_t)-2\,\mathrm{cut}_w(x,C,t)
}{
\min\!\bigl(\mathrm{vol}_w(C',t),\,\mathbf{W}_t-\mathrm{vol}_w(C',t)\bigr)
},
\end{equation}
where $\mathbf{W}_t=\sum_{u\in U} d_w(u,G_t)$ denotes the total wedge volume on the query side in $G_t$.
\end{corollary}

By Corollary \ref{eq:con-update}, the incremental computation of $\phi_w(C',t)$ only requires $\mathrm{cut}_w(x,C,t)$, which has already been maintained during the priority-driven expansion. Therefore, once a vertex $x$ is selected, the wedge conductance of the expanded community can be updated in constant time.

\myparagraph{Pruning Rules}
During the best-first expansion, the number of candidate supersets of $q$ grows combinatorially. To reduce the search space, we exploit the properties of the $(\alpha,\beta,\tau)$-wedge core and devise several pruning rules to eliminate unnecessary extensions.

\underline{\textit{Frequency-based Pruning.}}
The frequency constraint is an inherent requirement of the $(\alpha,\beta,\tau)$-wedge core. A key observation is that the candidate timestamp set of a community is monotone under expansion, which enables pruning based solely on temporal feasibility.

\begin{prule}
\label{lem:fre-mono}
Let $C\subseteq U$ be a candidate community, and let $\mathcal{CT}_C$ denote its candidate timestamp set. If $|\mathcal{CT}_C|<\tau$, then for any superset $C'\supseteq C$, we have $\mathcal{CT}_{C'}\subseteq \mathcal{CT}_C$, and thus $|\mathcal{CT}_{C'}|<\tau$. Therefore, $C$ and all its descendants can be safely pruned.
\end{prule}

\begin{prule}
\label{lem:fre-bound}
Let the ordered candidate timestamps of $C$ be $\mathcal{CT}_C=\{t_1,\dots,t_m\}$. Suppose the algorithm has processed the first $k$ timestamps in $\mathcal{CT}_C$, and let $f(C)$ denote the number of valid timestamps of $C$ found so far. If $f(C)+(m-k)<\tau,$ then even if all remaining timestamps become valid, the total number of valid timestamps of $C$ is still less than $\tau$. Hence, $C$ and all its descendants can be safely pruned.
\end{prule}

{{\underline{\textit{Conductance-based Pruning.}}}}
While Rules \ref{lem:fre-mono} and \ref{lem:fre-bound} remove temporally infeasible candidates, they do not exploit the quality of feasible communities. We therefore use the current global best solution to prune unnecessary computations and suboptimal branches.

\begin{prule}
\label{lem:con-lb}
Let $\Phi^*$ be the minimum temporal wedge conductance found so far. For a candidate community $C$, let $\Phi_k(C)$ denote the intermediate temporal wedge conductance computed from the valid timestamps identified so far in the current scan order over $\mathcal{CT}_C$. Since $\Phi_k(C)\le \Phi(C)$ always holds, if $\Phi_k(C)\ge \Phi^*$ at any step, then the exact computation of $\Phi(C)$ can be terminated early, because $C$ cannot improve the current best solution.
\end{prule}

\begin{lemma}
\label{lem:subtree-lb}
Let $C\subseteq U$ be a candidate community with candidate timestamp set $\mathcal{CT}_C$. Let $\mathcal{F}$ denote the set of vertices removed in Stage-1 filtering, so that every feasible descendant $C'\supseteq C$ satisfies $C'\cap \mathcal{F}=\emptyset$. Define the infeasible extension set of $C$ as $\mathcal{B}(C)=\{x\in U\setminus C \mid |\mathcal{CT}_C\cap \mathcal{CT}(x)|<\tau\}$, and let the overall invalid set be $\mathcal{I}(C)=\mathcal{B}(C)\cup \mathcal{F}$. For each $t\in \mathcal{CT}_C$, define
\begin{equation}
\underline{\mathrm{cut}}_w(C,t)=\sum_{v\in V}|N_t(v)\cap C|\cdot |N_t(v)\cap \mathcal{I}(C)|,
\label{eq:cutw}
\end{equation}
and let $\underline{\phi}(C,t)=\underline{\mathrm{cut}}_w(C,t)/(\mathbf{W}_t/2)=2\,\underline{\mathrm{cut}}_w(C,t)/\mathbf{W}_t$, where $\mathbf{W}_t=\sum_{u\in U} d_w(u,G_t)$ is the total wedge volume on the query side. Let $\underline{\phi}_{(1)}(C)\le \underline{\phi}_{(2)}(C)\le \cdots \le \underline{\phi}_{(m)}(C)$ be the nondecreasing rearrangement of $\{\underline{\phi}(C,t)\mid t\in \mathcal{CT}_C\}$, where $m=|\mathcal{CT}_C|$. Further define
\begin{equation}
   \mathrm{LB}(C)=\min_{\tau\le k\le m}\frac{1}{k}\sum_{i=1}^{k}\underline{\phi}_{(i)}(C).
\end{equation}
Then, for any feasible descendant $C'\supseteq C$ with $|\mathcal{CT}_{C'}|\ge\tau$, it holds that $\Phi(C')\ge \mathrm{LB}(C)$. Consequently, if $\mathrm{LB}(C)\ge \Phi^*$, the entire branch rooted at $C$ can be pruned safely.
\end{lemma}

For any vertex $u\in U^{cand}$, let $\overline{W}(u)$ denote the aggregated candidate wedge-neighbor set of $u$ over the query interval, i.e., $\overline{W}(u)=\{x\in U^{cand}\setminus\{u\}\mid \exists t\in \mathcal{T}_Q,\ x\in W_t(u,G_t)\}$. Accordingly, the candidate expansion set of a community is maintained by progressively aggregating the wedge neighbor candidates of its newly inserted vertices.

\begin{algorithm}[t]
\footnotesize
\SetKwInOut{Input}{Input}
\SetKwInOut{Output}{Output}
\SetKwFunction{ComputeMTS}{ComputeMTS}
\SetKwFunction{ComputeLB}{ComputeLB}
\SetKwFunction{CheckWCore}{CheckWCore}
\SetAlgoVlined
\caption{\textsf{Expansion} $(\mathcal{G}_{cand}, q, \mathcal{T}_Q, \alpha, \beta, \tau)$}
\label{alg:expansion}
\Input{$\mathcal{G}_{cand}=(U^{cand},V^{cand},\mathcal{E})$: candidate graph;\\
$\alpha,\beta,\tau$: constraints; $q,\mathcal{T}_Q$: query node and interval}
\Output{$C^*$: optimal $(\alpha,\beta,\tau)$-wedge core community}

$C_{init}\leftarrow\{q\},\ \mathcal{CT}_{C_{init}}\leftarrow \mathcal{T}_Q\cap \mathcal{CT}(q),\ C^*\leftarrow\{q\},\ \Phi^*\leftarrow +\infty$\;
$\forall t \in \mathcal{CT}_{C_{init}}:\ \mathrm{vol}_w(C_{init},t) \leftarrow d_w(q,G_t),\ \mathrm{cut}_w(C_{init},t) \leftarrow d_w(q,G_t)$\;
$Cand(C_{init})\leftarrow \overline{W}(q)$\;
$Q\leftarrow \emptyset$\;
$Q.\mathrm{push}(C_{init},\ \max_{x\in Cand(C_{init})}\mathcal{P}(x\mid C_{init}))$\;

\While{$Q\neq \emptyset$}{
    $C\leftarrow Q.\mathrm{pop}()$\;
    \ForEach{$u\in Cand(C)$}{
        $C'\leftarrow C\cup\{u\},\ \mathcal{CT}_{C'}\leftarrow \mathcal{CT}_C\cap \mathcal{CT}(u)$\;
        \lIf{$|\mathcal{CT}_{C'}|<\tau$}{\textbf{continue}}
        $\mathcal{T}_{C'}\leftarrow \emptyset,\ WCS[C']\leftarrow \emptyset,\ cnt\leftarrow 0,\ rem\leftarrow |\mathcal{CT}_{C'}|$\;

        \ForEach{$t\in \mathcal{CT}_{C'}$}{
            $rem\leftarrow rem-1$\;
            \If{$|C'|<\alpha+1$}{
                \lIf{$cnt+rem<\tau$}{\textbf{break}}
                \textbf{continue}\;
            }
            \If{\CheckWCore$(C,u,t)$}{
                $\mathcal{T}_{C'}\leftarrow \mathcal{T}_{C'}\cup\{t\},\ cnt\leftarrow cnt+1$\;
                $\mathrm{vol}_w(C',t)\leftarrow \mathrm{vol}_w(C,t)+d_w(u,G_t)$\;
                $\mathrm{cut}_w(C',t)\leftarrow \mathrm{cut}_w(C,t)+d_w(u,G_t)-2\,\mathrm{cut}_w(u,C,t)$\;
                $WCS[C'].\mathrm{append}\!\left(\frac{\mathrm{cut}_w(C',t)}{\min(\mathrm{vol}_w(C',t),\,\mathbf{W}_t-\mathrm{vol}_w(C',t))}\right)$\;
            }
            \lIf{$cnt+rem<\tau$}{\textbf{break}}
        }

        \lIf{$|\mathcal{T}_{C'}|<\tau$}{\textbf{continue}}
        $\Phi_{curr}\leftarrow \ComputeMTS(C',WCS[C'],\tau)$\;
        \If{$\Phi_{curr}<\Phi^*$}{
            $\Phi^*\leftarrow \Phi_{curr},\ C^*\leftarrow C'$\;
        }
        $\mathrm{LB}(C')\leftarrow \ComputeLB(C')$\;
        \If{$LB_{C'}<\Phi^*$}{
            $Cand(C')\leftarrow (Cand(C)\cup \overline{W}(u))\setminus C'$\;
            $Q.\mathrm{push}(C',\ \max_{x\in Cand(C')}\mathcal{P}(x\mid C'))$\;
        }
    }
}
\Return{$C^*$}\;
\end{algorithm}

The overall expansion procedure is detailed in Algorithm \ref{alg:expansion}. It first initializes the query-anchored community $C_{init}=\{q\}$ together with its candidate timestamps, wedge volume, wedge cut, candidate set, and priority queue (Lines 1--5). Then, the algorithm repeatedly extracts a community $C$ from the queue and explores its candidate wedge neighbors (Lines 6--30). For each possible expansion $C'=C\cup\{u\}$, it first intersects candidate timestamps and immediately prunes the branch if the frequency requirement is already violated (Lines 9--10). Next, it scans the timestamps in $\mathcal{CT}_{C'}$, maintains the number of unprocessed candidate timestamps, performs structural validation through \textit{CheckWCore}, and incrementally updates the wedge volume, wedge cut, and wedge conductance sequence $WCS[C']$ whenever $C'$ is valid at timestamp $t$ (Lines 12--22). If $C'$ is valid on fewer than $\tau$ timestamps, it is discarded (Line 23). Otherwise, the algorithm computes $\Phi_{curr}$ via Algorithm \ref{alg:MTS}, updates the current global optimum if necessary (Lines 24--26), and then evaluates the subtree lower bound $LB_{C'}$ via \textit{ComputeLB} according to Lemma \ref{lem:subtree-lb} (Line 27). The new state $C'$ is inserted into the queue only when this lower bound indicates that the branch still has the potential to improve the current best solution (Lines 28--30).

\begin{theorem}[Time Complexity Analysis]
\label{the:timecomplexity}
The time complexity of returning the optimal community via Algorithm \ref{alg:expansion} is $O\left(|\mathcal{S}| \cdot n \cdot |\mathcal{T}| \cdot d_{max}(U) \cdot d_{max}(V)\right)$, where $|\mathcal{S}|$ is the number of visited states, $n$ denotes the maximum candidate set size, and $d_{max}(U)$ and $d_{max}(V)$ represent the maximum degrees in $U$ and $V$, respectively.
\end{theorem}

\begin{example}
Fig. \ref{fig:expansion} provides a toy example to illustrate the expansion stage. Starting from $C_0=\{q\}=\{u_1\}$, the best-first rule ranks candidates by conductance-based priority and iteratively expands the current community by incorporating a wedge neighbor. As shown in Fig. \ref{fig:expansion}(c), the search reaches the feasible and low-conductance community $\{u_1,u_2,u_3\}$ early, updating the global optimum to $\Phi^*=1/7$. The tighter bound $\Phi^*$ further improves pruning effectiveness. Due to space limitations, only representative branches are depicted. The CT-frequency constraint removes entire descendants whose timestamp intersections are insufficient, e.g., $u_8$ with $|\mathcal{CT}_C\cap\mathcal{CT}(u_8)|<\tau$. At $C=\{u_1,u_2,u_4,u_6\}$, we have $LB(C)\approx 0.1742>\Phi^*$, and thus the whole subtree can be safely pruned according to Lemma \ref{lem:subtree-lb}. The other red nodes are excluded because they fail to satisfy the $(\alpha,\beta,\tau)$-wedge core constraint, e.g., $\{u_1,u_6,u_4\}$, or have suboptimal quality, e.g., $\{u_1,u_2,u_6\}$. In this running example, the algorithm finally returns $C^*=\{u_1,u_2,u_3\}$ with $\Phi^*=0.14$.
\end{example}

%% file: section/Method2.tex
\section{Compressed Offline Index}
\label{sec:method2}

The filtering stage of $PFCS$ becomes costly on large graphs. A naive snapshot-wise index is also inefficient, since it incurs prohibitive space complexity $O(|\mathcal{T}| \cdot \alpha_{\max} \cdot \beta_{\max} \cdot (|U| + |V|))$ due to structural and temporal redundancies, while its candidate set reconstruction still costs $O(|\mathcal{T}_Q|(|U|+|V|))$ time. To address this issue, we develop a compressed offline index for efficient query processing.

\subsection{Index Structure}
\label{sec:index-s}
Here, we propose a compressed index WCI (\underline{W}edge \underline{C}ore \underline{I}ndex) consisting of two parts, $\mathrm{WCI}^U$ and $\mathrm{WCI}^V$, for nodes in $U$ and $V$, respectively. Unlike snapshot-based indices, WCI partitions the vertex-time space via skyline coreness and aggregates temporal information into compressed patterns.

\begin{definition}[Skyline Coreness]
\label{de:skyline}
Given a snapshot $G_t$ and a vertex $w \in U \cup V$, its skyline coreness is defined as $\mathcal{K}_t(w) = \max_{\preceq}\{(\alpha,\beta)\mid w\in \mathcal{W}_{\alpha,\beta}(G_t)\}$, where $\max_{\preceq}$ returns the set of Pareto-maximal pairs under the order $(\alpha,\beta)\preceq(\alpha',\beta') \Leftrightarrow \alpha\le \alpha',\ \beta\le \beta'$.
\end{definition}

Based on Definition \ref{de:skyline}, $\mathrm{WCI}^U$ ($\mathrm{WCI}^V$) is composed of two main components, namely Pattern Bucket (PB) and Lookup Grid (LG), which are described as follows.

\myparagraph{Pattern Bucket (PB)} For each pair $(\alpha, \beta)$, we maintain exactly one bucket $\mathcal{PB}_{\alpha,\beta}$ to record all vertices $u \in U$ along with the corresponding timestamps during which they exhibit the skyline coreness $(\alpha, \beta)$. Observing that many such vertices share the same timestamp set, the bucket is organized as a linked list of pattern blocks, denoted as $\mathcal{P}_{\alpha, \beta}^{(1)} \rightarrow \mathcal{P}_{\alpha, \beta}^{(2)} \rightarrow \dots$. The $k$-th block is defined as a tuple $\mathcal{P}_{\alpha, \beta}^{(k)} = \langle TP_k, VP_k \rangle$, in which $VP_k$ is the list of vertices that share an identical timestamp pattern, and $TP_k \subseteq \mathcal{T}$ is the corresponding timestamp set. Specifically, for any vertex $u \in VP_k$, we have $TP_k=\{t\mid (\alpha,\beta)\in\mathcal{K}_t(u)\}$.

\myparagraph{Lookup Grid (LG)} The upper level of $\mathrm{WCI}^U$ is a 2D array covering the parameter space dimensions $\alpha_{\max} \times \beta_{\max}$. The entry $\mathrm{WCI}^U[\alpha][\beta]$ points to the first pattern block $\mathcal{P}_{\alpha, \beta}^{(1)}$. This entry is non-null if and only if there exists at least one vertex $u \in U$ and a timestamp $t$ such that $(\alpha, \beta) \in \mathcal{K}_t(u)$.

To facilitate rapid $\mathcal{CT}$ reconstruction without exhaustive grid scanning, we maintain an auxiliary directory $\mathcal{D}$, where $\mathcal{D}[u]$ stores pointers to all pattern blocks containing vertex $u$. Given a query $Q(q,\alpha,\beta,\tau,[t_s,t_e])$, we first traverse $\mathcal{D}[q]$ to collect the blocks satisfying $(\alpha',\beta') \succeq (\alpha,\beta)$, and reconstruct $\mathcal{CT}(q)$ by taking the union of their timestamp sets within $[t_s,t_e]$. If $|\mathcal{CT}(q)| < \tau$, the query terminates immediately. Otherwise, we scan the non-empty entries in the lookup grids of both $\mathrm{WCI}^U$ and $\mathrm{WCI}^V$ that satisfy $(\alpha', \beta') \succeq (\alpha, \beta)$. For each accessed block $\mathcal{P}_{\alpha', \beta'}^{(k)} = \langle TP_k, VP_k \rangle$, we merge the valid intersection $TP_k \cap \mathcal{CT}(q)$ into the timestamp set of each vertex $v \in VP_k$. Finally, vertices covering at least $\tau$ distinct timestamps form the candidate set.

\begin{figure*}[t]
\centering
\includegraphics[width=0.98\textwidth]{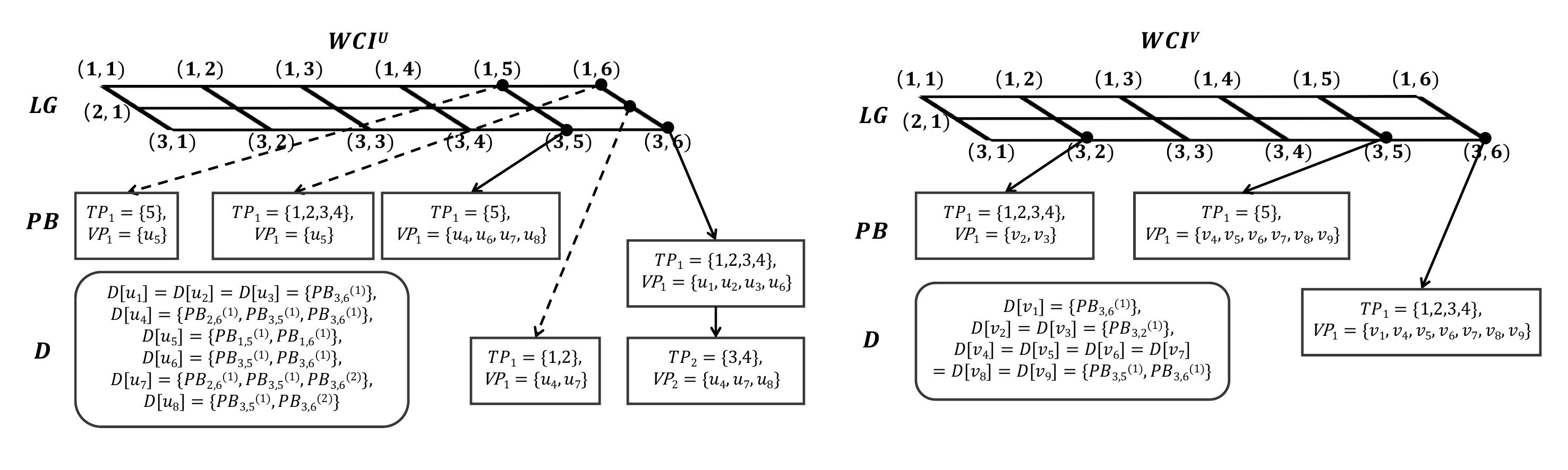}
\caption{The WCI built on the temporal bipartite graph in Figure \ref{fig:expansion}(a).}
\label{fig:index}
\end{figure*}

\begin{example}
Figure \ref{fig:index} illustrates the WCI built on Figure \ref{fig:expansion}(a). For the query $Q(q=u_1,\alpha=2,\beta=2,\tau=3,[t_1,t_5])$, WCI first traverses $\mathcal{D}[u_1]$ and reconstructs the candidate timestamp set $\mathcal{CT}(u_1)=\{1,2,3,4\}$. It then scans all non-empty skyline cells satisfying $(\alpha',\beta')\succeq(2,2)$ and accumulates the candidate timestamps of the accessed pattern blocks intersecting $\mathcal{CT}(u_1)$. Although $u_8$ is indexed at $(3,6)$ on $\{3,4\}$ and at $(3,5)$ on $\{5\}$, the latter contributes nothing since $\{5\}\cap\mathcal{CT}(u_1)=\emptyset$, leaving $u_8$ with only two candidate timestamps and thus failing $\tau=3$. Finally, $\mathrm{WCI}^U$ returns $U_{cand}=\{u_1,u_2,u_3,u_4,u_6,u_7\}$, while the symmetric scan on $\mathrm{WCI}^V$ yields $V_{cand}=V$. The induced candidate subgraph is then passed to the subsequent best-first expansion.
\end{example}

\subsection{Index Construction}
\label{sec:index-c}
In this subsection, we present an efficient $WCI$ construction algorithm. The overall framework is shown in Algorithm \ref{alg:WCICon}, while the skyline computation on each snapshot is detailed in Algorithm \ref{alg:Skyline}. Since $\mathrm{WCI}^V$ is constructed similarly to $\mathrm{WCI}^U$, we focus on $\mathrm{WCI}^U$ for exposition.

Different from standard $(\alpha,\beta)$-core decomposition \cite{liu2020efficient}, Algorithm \ref{alg:Skyline} adopts a wedge-based generalized peeling strategy to compute the skyline coreness sets on a snapshot $G_t$. It starts by initializing the active vertex set $S$, the skyline sets $\mathcal{K}_t(\cdot)$, and the common neighbor counters $\mathrm{coU}$ and $\mathrm{coV}$ (Lines 1--4). Then, it identifies the bottleneck pair $(\alpha,\beta)$ and collects into a queue $Q$ all vertices whose wedge neighbor numbers attain the corresponding minimum thresholds (Lines 6--7). Subsequently, the algorithm performs an iterative peeling procedure (Lines 8--13). Each time a vertex $x$ is removed, the pair $(\alpha,\beta)$ is inserted into $\mathcal{K}_t(x)$ unless it is dominated by an existing entry. Meanwhile, the deletion of $x$ triggers incremental updates to the maintained common neighbor counters and wedge neighbors. When $x\in U$, the update propagates first to the affected vertices in $V$ through the common neighbor counters induced by $N(x,S)$, and then to the relevant vertices in $U$ via the adjacent hub vertices (Lines 14--27). The case for $x\in V$ is handled symmetrically. Repeating this process until $S=\emptyset$ yields all skyline coreness pairs of the vertices in $G_t$.

\begin{algorithm}[t]
\caption{ComSkyline}
\label{alg:Skyline}
\footnotesize
\SetKwInOut{Input}{Input}
\SetKwInOut{Output}{Output}
\Input{$G_t=(U,V,E_t)$}
\Output{$\{\mathcal{K}_t(u)\}_{u\in U},\ \{\mathcal{K}_t(v)\}_{v\in V}$}

$S \leftarrow U \cup V$; $\forall x\in S,\ \mathcal{K}_t(x)\leftarrow \emptyset$\;
$\forall u<u'\in U,\ \mathrm{coU}[\{u,u'\}] \leftarrow |N(u,S)\cap N(u',S)|$\;
$\forall v<v'\in V,\ \mathrm{coV}[\{v,v'\}] \leftarrow |N(v,S)\cap N(v',S)|$\;
Compute $W(x,S)$ for all $x\in U\cup V$ from $\mathrm{coU}$ and $\mathrm{coV}$\;

\While{$S\neq\emptyset$}{
  $(\alpha,\beta) \leftarrow \big(\min_{u\in S\cap U}|W(u,S)|,\ \min_{v\in S\cap V}|W(v,S)|\big)$\;
  $Q \leftarrow \{u\in S\cap U\mid |W(u,S)|=\alpha\}\ \cup\ \{v\in S\cap V\mid |W(v,S)|=\beta\}$\;
  \While{$Q\neq\emptyset$}{
    $x \leftarrow \text{head}(Q)$; $Q \leftarrow Q \setminus \{x\}$\;
    \lIf{$x\notin S$}{\textbf{continue}}
    \If{$\nexists(\alpha',\beta')\in\mathcal{K}_t(x)\ \text{s.t.}\ (\alpha',\beta')\succeq(\alpha,\beta)$}{
      $\mathcal{K}_t(x)\leftarrow\mathcal{K}_t(x)\cup\{(\alpha,\beta)\}$\;
    }
    $S \leftarrow S \setminus \{x\}$\;
    \If{$x\in U$}{
      \ForEach{$v_i<v_j \in N(x,S)$}{
        $\mathrm{coV}[\{v_i,v_j\}] \leftarrow \mathrm{coV}[\{v_i,v_j\}] - 1$\;
        \If{$\mathrm{coV}[\{v_i,v_j\}]=0$}{
          $W(v_i,S)\leftarrow W(v_i,S)\setminus\{v_j\}$\;
          $W(v_j,S)\leftarrow W(v_j,S)\setminus\{v_i\}$\;
          \lIf{$|W(v_i,S)|<\beta$}{$Q \leftarrow Q \cup \{v_i\}$}
          \lIf{$|W(v_j,S)|<\beta$}{$Q \leftarrow Q \cup \{v_j\}$}
        }
      }
      \ForEach{$v\in N(x,S)$}{
        \ForEach{$u'\in N(v,S)\cap U,\ u'\neq x$}{
          $\mathrm{coU}[\{x,u'\}] \leftarrow \mathrm{coU}[\{x,u'\}] - 1$\;
          \If{$\mathrm{coU}[\{x,u'\}]=0$}{
            $W(u',S)\leftarrow W(u',S)\setminus\{x\}$\;
            \lIf{$|W(u',S)|<\alpha$}{$Q \leftarrow Q \cup \{u'\}$}
          }
        }
      }
    }
    \If{$x\in V$}{
      symmetrically update by exchanging $U$ and $V$, $\alpha$ and $\beta$, and $\mathrm{coU}$ and $\mathrm{coV}$ (Lines 14--27)\;
    }
  }
}
\Return $\{\mathcal{K}_t(u)\}_{u\in U},\ \{\mathcal{K}_t(v)\}_{v\in V}$\;
\end{algorithm}

\begin{algorithm}[t]
\caption{WCICon}
\label{alg:WCICon}
\footnotesize
\SetKwInOut{Input}{Input}
\SetKwInOut{Output}{Output}
\Input{temporal bipartite graph $\mathcal{G}=\{G_t\mid t\in\mathcal{T}\}$}
\Output{Index $\mathrm{WCI}$ ($\mathrm{WCI}^U$, $\mathrm{WCI}^V$) and directory $\mathcal{D}$}

$\mathcal{M}\leftarrow \emptyset$; $\mathrm{WCI}^U\leftarrow \textsf{null}$; $\mathcal{D}\leftarrow \emptyset$\;

\ForEach{$t\in\mathcal{T}$}{
  $\{\mathcal{K}_t(u)\}_{u\in U}, \{\mathcal{K}_t(v)\}_{v\in V} \leftarrow \textsf{ComSkyline}(G_t)$\;
  \ForEach{$u\in U, (\alpha,\beta)\in \mathcal{K}_t(u)$}{
     $\mathcal{M}[u,\alpha,\beta]\leftarrow \mathcal{M}[u,\alpha,\beta]\cup\{t\}$\;
  }
}
\ForEach{$(u,\alpha,\beta)\in \mathrm{dom}(\mathcal{M})$}{
$TP \leftarrow \mathcal{M}[u,\alpha,\beta]$\;
\If{$\exists \mathcal{P} \in \mathcal{PB}_{\alpha,\beta} \text{ s.t. } \mathcal{P}.TP = TP$}{
     $\mathcal{P}.VP \leftarrow \mathcal{P}.VP \cup \{u\}$\;
  }
\Else{
$\mathcal{P} \leftarrow \langle TP, \{u\} \rangle$\;
$\mathcal{PB}_{\alpha,\beta} \leftarrow \mathcal{PB}_{\alpha,\beta} \cup \{\mathcal{P}\}$\;
\lIf{$\mathrm{WCI}^U[\alpha][\beta] = \textsf{null}$}{$\mathrm{WCI}^U[\alpha][\beta]\leftarrow \textsf{ptr}(\mathcal{P})$}
}
$\mathcal{D}[u]\leftarrow \mathcal{D}[u]\cup\{\textsf{ptr}(\mathcal{P})\}$\;
}
Construct $\mathrm{WCI}^V$ and update $\mathcal{D}$ for $V$ symmetrically (Lines 4--14)\;
\Return{$\mathrm{WCI} = \{\mathrm{WCI}^U, \mathrm{WCI}^V\}, \mathcal{D}$}\;
\end{algorithm}

Furthermore, Algorithm \ref{alg:WCICon} follows a scan-and-materialize paradigm to construct the index. Specifically, we invoke Algorithm \ref{alg:Skyline} to compute the skyline coreness sets $\mathcal{K}_t(w)$ of all vertices on $G_t$ for each $t \in \mathcal{T}$ (Lines 2--5). For each occurrence of $(\alpha,\beta)\in\mathcal{K}_t(u)$ where $u \in U$, the timestamp $t$ is recorded into a temporary accumulator keyed by the triple $(u,\alpha,\beta)$ (Lines 4--5). After scanning all snapshots, each triple is associated with an aggregated timestamp set indicating when vertex $u$ possesses the skyline pair $(\alpha,\beta)$. The pattern buckets and lookup grid are then materialized by compressing these temporal records (Lines 6--15). For each triple $(u,\alpha,\beta)$, the corresponding timestamp set $TP$ is inserted into the bucket $\mathcal{PB}_{\alpha,\beta}$. If there already exists a pattern block with the same $TP$, then $u$ is merged into its vertex list. Otherwise, a new block $\langle TP,\{u\}\rangle$ is created (Lines 8--12). To ensure storage compactness, the grid entry $\mathrm{WCI}^U[\alpha][\beta]$ is initialized to point to a new pattern block only upon the first occurrence of $(\alpha,\beta)$ (Line 12). Simultaneously, the vertex directory $\mathcal{D}$ is updated with pointers to the respective blocks, mapping each vertex $u$ to its compressed patterns (Line 13). The computational complexity of Algorithm \ref{alg:WCICon} consists of two parts: the snapshot-wise skyline computation by Algorithm \ref{alg:Skyline} and the subsequent materialization of the recorded skyline memberships, detailed in Theorem \ref{the:WCI}.

\begin{theorem}
\label{the:WCI}
The time complexity of Algorithm \ref{alg:WCICon} is $O(\Delta\cdot m)$, where $m=\sum_{t\in\mathcal{T}}|E_t|$, $\Delta=\max_{t\in\mathcal{T}}\max_{x\in U\cup V} d_t(x)$, and $d_t(x)=|N_t(x)|$ denotes the degree of $x$ in $G_t$.
\end{theorem}

%% file: section/Experiment.tex
\section{Experiment}
\label{exp}
In this section, we conduct extensive experiments to evaluate the effectiveness and efficiency of our solutions. All algorithms are implemented in C++ and compiled with g++ 11.4.0 using -O3 optimization. All experiments are conducted on a Linux machine with an Intel Xeon(R) Silver 4210@2.20GHz CPU and 1TB RAM.

\subsection{Experimental Setup}

\myparagraph{Datasets}
We evaluate our solutions on seven widely used datasets (Table \ref{tab:dataset}). 
Ipvevents\footnote{https://tianchi.aliyun.com/dataset/123862} (Ip) is a real customer-product network, where edges denote clicking interactions between customers and products. The remaining six datasets are publicly available from KONECT.\footnote{http://konect.cc/networks/} In particular, \textit{diq}, \textit{vec} and \textit{ar} are Wikipedia edit networks, \textit{LK} is the Linux kernel mailing list reply network, \textit{Wut} is a Twitter user-tag network, and \textit{Bti} is a BibSonomy tag-item network. 

\myparagraph{Algorithms}
The following algorithms are implemented for comparison. (1) Static bipartite community search methods: \textit{SCC} \cite{ZHOUSize23}, \textit{SABC} \cite{zhang2024size} and \textit{Top-r} \cite{zhang2150top}. These methods are originally designed for static bipartite graphs. To adapt them to our setting, we first aggregate the temporal bipartite graph over the query interval into a static bipartite graph and then run the methods on it. \textit{SCC} and \textit{SABC} search for query-centered cohesive communities under size constraints based on $(\alpha, \beta)$-core, whereas \textit{Top-r} addresses top-$r$ influential community search rather than query-centered community search. Therefore, for \textit{Top-r}, we use the community containing the query vertex for evaluation. (2) Temporal unipartite community search methods: \textit{QTCS} \cite{LinQTCS24} and \textit{MCTS} \cite{Zhong25kcore}. \textit{QTCS} is a query-centered temporal community search method based on temporal proximity, whereas \textit{MCTS} returns the temporal $k$-core connected component containing the query vertex over the specified time interval. To adapt these two methods to our setting, we first project the temporal bipartite graph onto the query side, execute the methods on the projected graph, and then map the results back for evaluation. (3) Temporal bipartite community search methods: \textit{PCSearch} \cite{li2023persistent}, \textit{RCSearch} \cite{LiMoreliable} and \textit{TABC} \cite{tabc24}. \textit{PCSearch} searches for query-dependent persistent $(\alpha,\beta)$-core communities, while \textit{RCSearch} focuses on reliable $(\alpha,\beta)$-communities. \textit{TABC} retrieves structurally feasible communities over the query interval, but it is not originally designed for query-centered community search. Therefore, we use the connected component containing the query vertex from its output for evaluation. In addition, the following three algorithms for WCCS are also implemented and evaluated, namely \textit{BK}, \textit{PFCS} proposed in Section \ref{sec:method1}, and \textit{WCI} proposed in Section \ref{sec:method2}.

\begin{table}[t]\centering
  \caption{Datasets statistics.}
  \label{tab:dataset}
  \begin{tabular}{lccccc}
    \toprule
    \textbf{Dataset} & \textbf{$|U|$} & \textbf{$|V|$} & \textbf{$|\mathcal{E}|$}  & \textbf{$|\mathcal{T}|$}  \\
    \midrule
    Ip  & 28,540   & 37,088   & 73,153       & 31 \\
    diq & 25,771   & 1,526    & 133,874      & 12 \\
    vec & 33,587   & 2,282    & 339,722       & 14  \\
    LK  & 337,510  & 42,046   & 605,642       & 35  \\
    Wut & 530,419  & 175,215  & 2,118,877     & 39  \\
    Bti & 767,448  & 204,674  & 2,517,857    & 22 \\
    ar  & 2,943,712 & 209,374 & 13,601,759  & 57  \\
    \bottomrule
  \end{tabular}
\end{table}

\myparagraph{Parameters}
We conduct experiments by varying the three parameters $\alpha$, $\beta$, and $\tau$, whose default values are $(3, 3, 3)$. Unless otherwise specified, we fix two parameters at their default values and vary the remaining one to evaluate its effect. For each setting, we randomly generate 10 query instances and report the average results. 

\begin{figure}[t]
	\centering
	\includegraphics[width=0.95\linewidth]{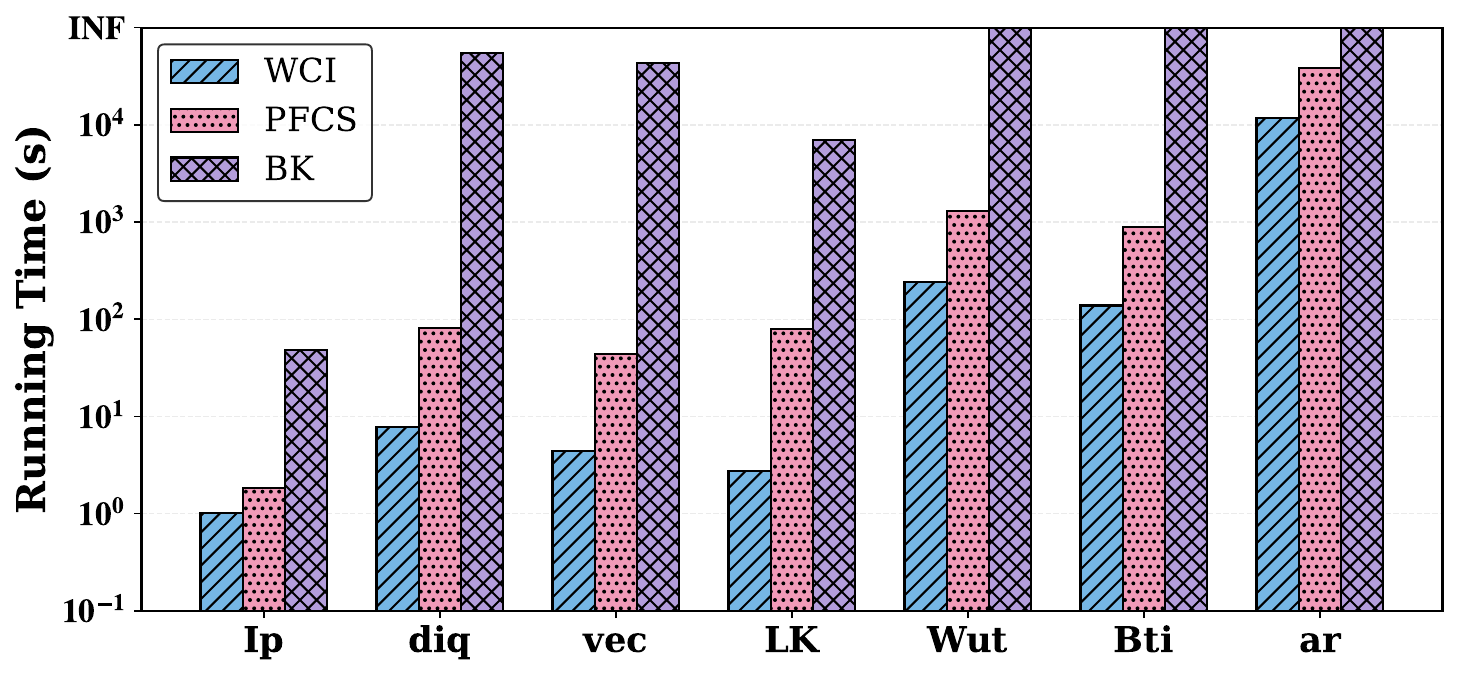}
	\caption{Response time on all the datasets}
	\label{fig:exp1}
\end{figure}

\subsection{Efficiency Evaluation}
Due to the objective functions of \textit{SCC}, \textit{SABC}, \textit{Top-r}, \textit{QTCS}, \textit{MCTS}, \textit{PCSearch}, \textit{RCSearch} and \textit{TABC} being different from the temporal wedge conductance studied in this work, directly comparing their query efficiency with our algorithms would be inappropriate. Therefore, the efficiency evaluation in this subsection focuses on our three algorithms, namely \textit{BK}, \textit{PFCS}, and \textit{WCI}. For those experiments that cannot finish within 24 hours, we mark them as INF. 

\myparagraph{Exp-1: Running time of different WCCS algorithms over all the datasets}
Figure \ref{fig:exp1} reports the query response time of \textit{BK}, \textit{PFCS}, and \textit{WCI} on all datasets. As can be seen, we have: (1) \textit{PFCS} consistently outperforms \textit{BK}, and the advantage becomes more pronounced on larger datasets. In particular, \textit{PFCS} is more than one order of magnitude faster than \textit{BK} on \textit{diq}, \textit{vec}, and \textit{LK}. Moreover, \textit{BK} fails to finish on \textit{Wut}, \textit{Bti}, and \textit{ar}, whereas \textit{PFCS} still returns results. (2) \textit{WCI} achieves the best performance on all datasets. Compared with \textit{PFCS}, it reduces the running time by about $1.8\times$ to $29.1\times$, showing the effectiveness of the proposed index-based acceleration strategy. These results clearly demonstrate the efficiency of our online optimization and index-based acceleration strategies.

\begin{figure}[t]
    \centering
    \includegraphics[width=0.49\columnwidth]{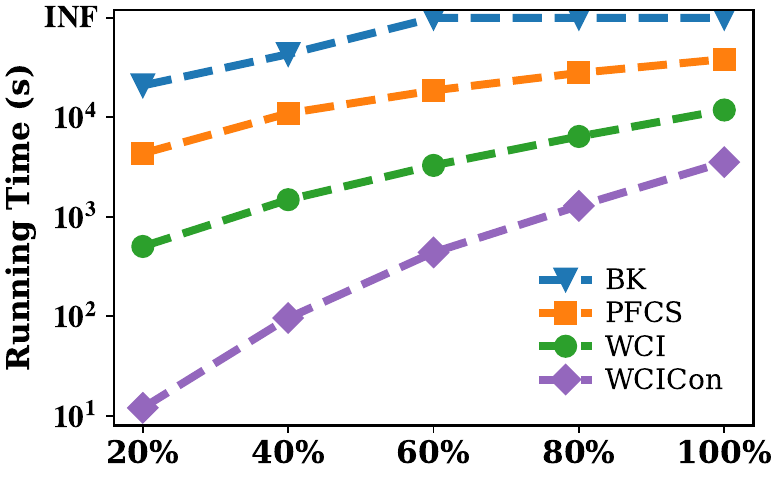}
    \hfill
    \includegraphics[width=0.49\columnwidth]{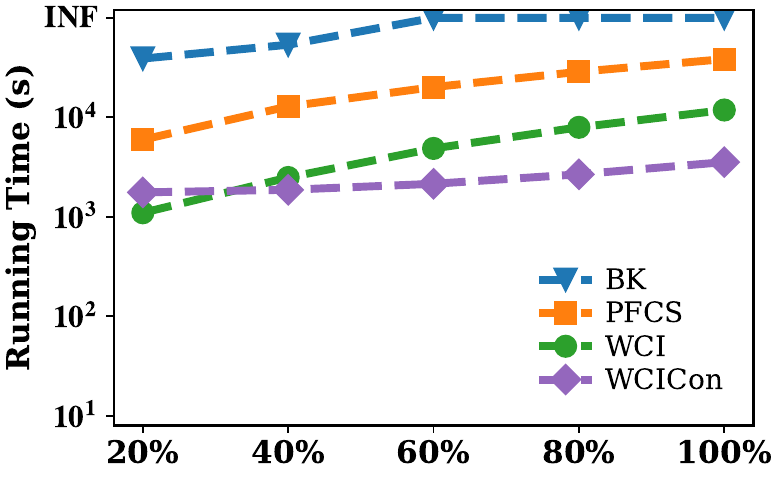}
    \vspace{-0.3em}
    \makebox[0.49\columnwidth][c]{(a) sampling $n$}
    \hfill
    \makebox[0.49\columnwidth][c]{(b) sampling $m$}
    \vspace{-0.4em}
    \caption{Scalability testing.}
    \label{fig:exp3}
\end{figure}

\myparagraph{Exp-2: Scalability testing}
In this experiment, we use the largest dataset \textit{ar} to demonstrate the scalability of \textit{BK}, \textit{PFCS}, \textit{WCI}, and the index construction algorithm \textit{WCICon} (Algorithm \ref{alg:WCICon}). Specifically, we randomly sample 20\% to 100\% of vertices and edges, corresponding to the two settings of sampling $n$ and sampling $m$ in Figure \ref{fig:exp3}(a) and \ref{fig:exp3}(b), respectively, to generate subgraphs of different sizes. As shown in Figure \ref{fig:exp3}, the running times of all methods increase as the graph size grows. Among them, \textit{BK} performs the worst and fails to return results once the graph becomes moderately large. In contrast, both \textit{PFCS} and \textit{WCI} increase smoothly, demonstrating favorable scalability under both sampling $n$ and sampling $m$. We also examine \textit{WCICon} separately. Its construction time increases with the graph size but remains well controlled, as shown in Figure \ref{fig:exp3}. This further confirms the scalability of the proposed indexing framework, since it measures the offline index construction cost rather than the online query time.

\begin{figure}[t]
	\centering
	\includegraphics[width=0.95\linewidth]{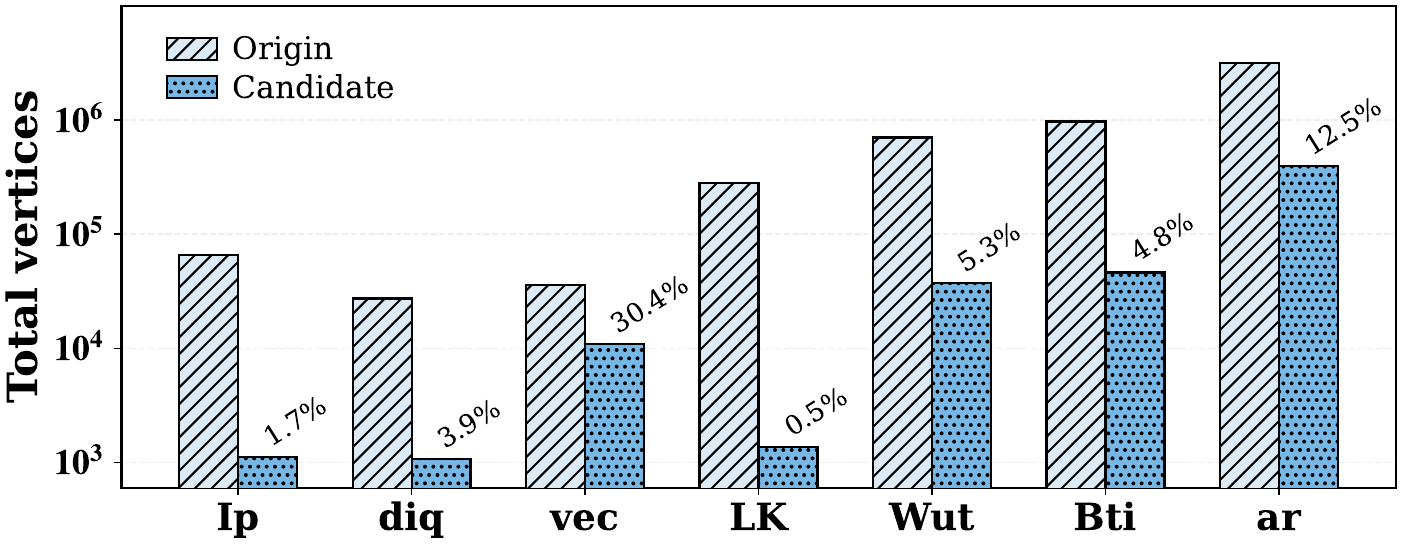}
	\caption{Evaluation of candidate set filter technique}
	\label{fig:exp4}
\end{figure}

\myparagraph{Exp-3: Evaluation of candidate set filtering technique}
To evaluate the candidate set filtering technique, we compare the total number of vertices in the original graph and in the filtered candidate graph for each dataset and report the remaining ratio of candidate vertices over original vertices in Figure \ref{fig:exp4}. As shown in the figure, the proposed filtering strategy is highly effective, as the candidate graph is consistently much smaller than the original graph on all datasets, which indicates that the filtering step can significantly reduce the search space before the subsequent query processing. In particular, the candidate graph only preserves a very small fraction of vertices on most datasets, e.g., about 1.7\% on \textit{Ip}, 3.9\% on \textit{diq}, and only 0.5\% on \textit{LK}. Even on the largest dataset \textit{ar}, the candidate graph contains only $395,176$ vertices compared with $3,153,086$ vertices in the original graph.

\begin{figure}[t]
	\centering
	\includegraphics[width=0.95\linewidth]{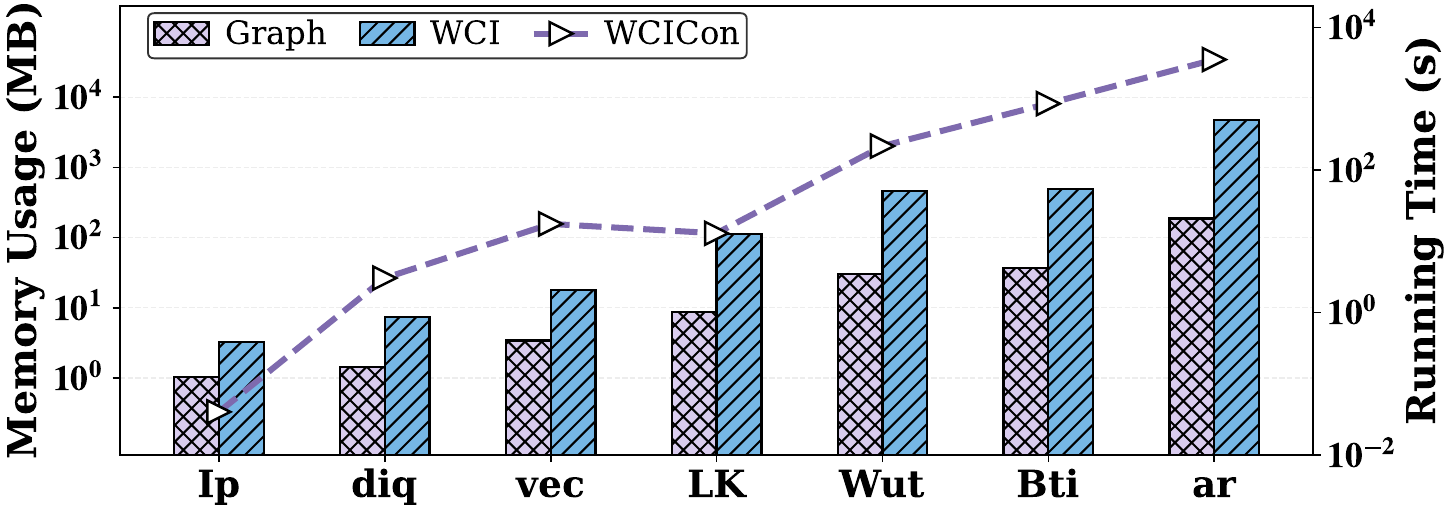}
	\caption{Graph size, index size, and construction time of \textit{WCI}.}
	\label{fig:exp5}
\end{figure}

\myparagraph{Exp-4: Evaluation of index construction time and index size}
Figure \ref{fig:exp5} reports the graph size, the index size of \textit{WCI}, and the construction time of \textit{WCICon} on all tested datasets. As shown in Figure \ref{fig:exp5}, the proposed indexing framework scales well in both space and time. We observed that the size of \textit{WCI} is about $3.1\times$ to $25.2\times$ the graph size on the tested datasets. For example, on \textit{Ip}, the graph and index sizes are 1.03 MB and 3.24 MB, respectively, while on the largest dataset \textit{ar}, they are 186.71 MB and 4710.04 MB. This behavior is consistent with our index design. Since \textit{WCI} explicitly organizes structural and temporal information for efficient query answering, it naturally requires more memory than the raw graph. However, it can still be maintained successfully for all datasets. A similar trend can be observed for the construction time of \textit{WCICon}. The construction cost increases with the dataset size, ranging from only 0.04 s on \textit{Ip} to 3545.62 s on \textit{ar}. Nevertheless, the index can still be constructed successfully within an acceptable time for all datasets. More importantly, this is a one-time offline cost, while the constructed index can be reused to support subsequent queries efficiently.

\begin{figure}[t]
    \centering
    \includegraphics[width=0.49\columnwidth]{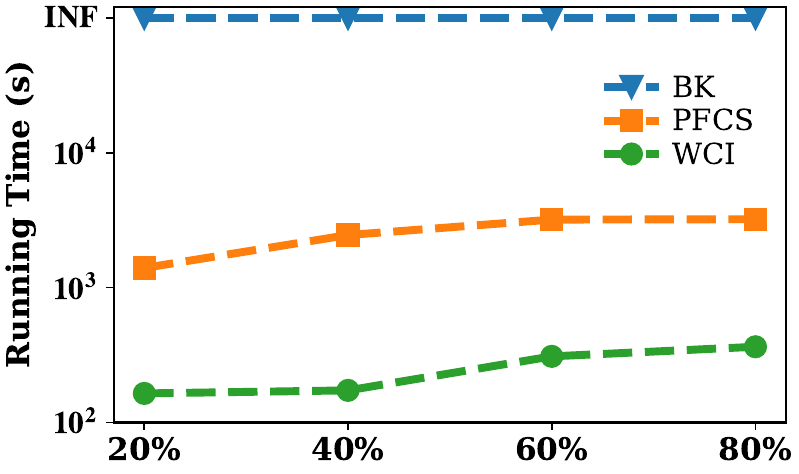}
    \hfill
    \includegraphics[width=0.49\columnwidth]{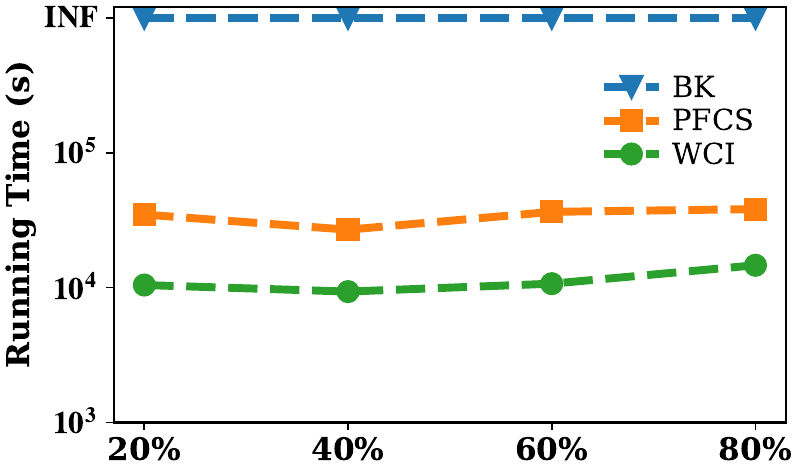}
    \vspace{-0.3em}
    \makebox[0.49\columnwidth][c]{(a) \textit{Wut} (varying $|\mathcal{T}_Q|$)}
    \hfill
    \makebox[0.49\columnwidth][c]{(b) \textit{ar} (varying $|\mathcal{T}_Q|$)}
    \vspace{-0.4em}
    \caption{Evaluation of query window size on \textit{Wut} and \textit{ar}.}
    \label{fig:exp6}
\end{figure}

\myparagraph{Exp-5: Evaluating the effect of query window size}
To evaluate the effect of query window size, we vary the query interval length from 20\% to 80\% of the whole time range on \textit{Wut} and \textit{ar}. As shown in Figure \ref{fig:exp6}, the response time generally increases with the window size for both \textit{PFCS} and \textit{WCI}, since a larger window introduces more timestamps and a larger temporal search space. Nevertheless, the increase remains smooth overall, which indicates that both methods, especially \textit{WCI}, are not overly sensitive to the growth of the temporal range. Meanwhile, \textit{WCI} consistently outperforms \textit{PFCS} under all settings, and its advantage remains stable as the window size increases. For example, when the window size reaches 80\%, \textit{WCI} is still about 8.8$\times$ faster than \textit{PFCS} on \textit{Wut} and about 2.6$\times$ faster on \textit{ar}. In contrast, \textit{BK} fails to return results on both datasets for all window sizes.

%% file: section/NewEffectiveness.tex
\subsection{Effectiveness Evaluation}
Evaluating the quality of temporal community search is challenging due to the absence of ground-truth communities in temporal bipartite networks. Following the common practice in temporal clustering and temporal community analysis, we evaluate the effectiveness of all methods from both auxiliary structural metrics and the objective metric perspective \cite{spi2004, LinQTCS24}. Specifically, we report temporal density (\textit{TD}), temporal conductance (\textit{TC}), and the proposed temporal wedge conductance (\textit{TWC}). Given a returned community $C\subseteq U$, let $G_t^U=(U,E_t^U)$ denote the $U$-side projected graph of snapshot $G_t$ for each $t\in\mathcal{T}_Q$. Then $TD(C)=\frac{1}{|\mathcal{T}_Q|}\sum_{t\in\mathcal{T}_Q}\frac{|\{(u,v)\in E_t^U \mid u,v\in C\}|}{\binom{|C|}{2}}$, and $TC(C)=\frac{\sum_{t\in\mathcal{T}_Q} |\{(u,v)\in E_t^U \mid u\in C,\ v\in U\setminus C\}|}{\sum_{t\in\mathcal{T}_Q}\sum_{u\in C} d_t^U(u)}$. A larger \textit{TD} and a smaller \textit{TC} indicate better structural quality. \textit{TWC} is computed on the original temporal bipartite graph under wedge semantics based on Definition \ref{twc}, and a smaller \textit{TWC} indicates better community quality.

\begin{table*}[htbp]
\centering
\caption{Community quality of different methods on all datasets. The best and second-best results in each metric are marked in \textbf{bold} and \underline{underlined}, respectively. Smaller TWC and TC indicate higher community quality, while larger TD indicates higher community quality. A baseline is excluded if it cannot finish within one day.}
\label{tab:effectiveness}
\begin{adjustbox}{width=0.95\textwidth,center}
\begin{tabular}{l!{\vrule width 0.8pt}ccccccc}
\toprule
TWC / TC / TD & \textit{Ip} & \textit{diq} & \textit{vec} & \textit{LK} & \textit{Wut} & \textit{Bti} & \textit{ar} \\
\midrule

\textit{QTCS}
& \underline{0.21} / 0.34 / \textbf{0.67}
& 0.68 / 0.58 / 0.31
& 0.84 / \textbf{0.14} / 0.53
& 0.46 / 0.36 / 0.49
& \textbf{0.45} / 0.74 / 0.31
& 0.96 / 0.88 / 0.34
& 0.89 / 0.86 / 0.20 \\

\textit{MCTS}
& 0.72 / 0.35 / \underline{0.65}
& 0.69 / \underline{0.52} / 0.42
& 0.54 / 0.48 / \underline{0.55}
& 0.70 / 0.41 / 0.47
& 0.57 / \underline{0.36} / 0.49
& \underline{0.62} / \textbf{0.34} / \underline{0.57}
& 0.86 / \textbf{0.31} / \underline{0.58} \\

\textit{SCC}
& 0.61 / 0.40 / 0.60
& \underline{0.39} / 0.87 / 0.26
& 0.72 / 0.55 / 0.47
& 0.59 / 0.45 / 0.43
& 0.90 / 0.98 / 0.12
& 0.66 / 0.89 / 0.22
& OOT \\

\textit{SABC}
& 0.58 / 0.37 / 0.57
& 0.42 / 0.88 / 0.32
& 0.68 / 0.57 / 0.41
& 0.63 / 0.48 / 0.51
& 0.82 / 0.79 / 0.19
& 0.75 / 0.84 / 0.38
& 0.77 / 0.64 / \textbf{0.59} \\

\textit{Top-r}
& 0.57 / 0.36 / 0.16
& 0.85 / 0.78 / 0.07
& 0.77 / \underline{0.44} / 0.18
& 0.82 / 0.79 / 0.09
& 0.79 / 0.63 / 0.14
& 0.90 / 0.85 / 0.05
& 0.93 / 0.81 / 0.17 \\

\textit{PCSearch}
& 0.35 / \textbf{0.18} / 0.46
& \textbf{0.22} / \textbf{0.38} / \textbf{0.67}
& \textbf{0.29} / \underline{0.44} / 0.45
& \underline{0.41} / \underline{0.35} / \textbf{0.70}
& OOT
& OOT
& OOT \\

\textit{RCSearch}
& 0.62 / 0.55 / 0.63
& 0.74 / 0.58 / 0.41
& 0.86 / 0.48 / 0.39
& 0.68 / 0.66 / \underline{0.53}
& 0.73 / 0.41 / \underline{0.51}
& 0.81 / 0.62 / 0.22
& \underline{0.72} / 0.83 / 0.18 \\

\textit{TABC}
& 0.63 / 0.42 / 0.27
& 0.57 / \textbf{0.38} / 0.13
& 0.74 / 0.61 / 0.26
& 0.61 / 0.39 / 0.22
& 0.77 / \textbf{0.27} / 0.46
& 0.84 / 0.67 / 0.49
& 0.81 / \underline{0.40} / 0.51 \\

\textit{Ours}
& \textbf{0.09} / \underline{0.32} / 0.59
& \textbf{0.22} / 0.58 / \underline{0.57}
& \underline{0.35} / 0.53 / \textbf{0.61}
& \textbf{0.07} / \textbf{0.33} / 0.41
& \underline{0.46} / 0.59 / \textbf{0.55}
& \textbf{0.49} / \underline{0.52} / \textbf{0.66}
& \textbf{0.61} / 0.57 / \underline{0.58} \\
\bottomrule
\end{tabular}
\end{adjustbox}
\end{table*}

\myparagraph{Exp-6: Community quality comparison between algorithms}
Table~\ref{tab:effectiveness} reports the community quality of all methods in terms of \textit{TWC}, \textit{TC}, and \textit{TD} on the seven datasets. Since some baselines cannot finish within the time limit on larger datasets, their missing results are marked as "OOT" in the table. For \textit{TWC}, our method achieves the best performance on most datasets, obtaining the best results on \textit{Ip}, \textit{diq}, \textit{LK}, \textit{Bti}, and \textit{ar}, and the second-best results on \textit{vec} and \textit{Wut}. This indicates that our method is more effective in finding communities with lower temporal wedge conductance. \textit{PCSearch} remains competitive on a few datasets such as \textit{diq} and \textit{vec}, but fails on larger datasets, which limits its applicability in practice. The remaining baselines generally perform worse, since temporal unipartite methods ignore the bipartite interaction pattern, static bipartite methods ignore temporal evolution, and \textit{TABC} is designed for temporal core retrieval rather than directly optimizing community quality.

For \textit{TC} and \textit{TD}, our method exhibits the most balanced overall performance, avoiding the common trade-off between the two metrics. In particular, it obtains the best \textit{TD} values on \textit{vec}, \textit{Wut}, and \textit{Bti}, and the second-best values on \textit{diq} and \textit{ar}, while remaining competitive on \textit{TC}. Although \textit{QTCS} and \textit{MCTS} often achieve relatively good \textit{TC} values due to their conductance-related objectives, their \textit{TD} values are usually much worse than ours, indicating that they cannot balance temporal separateness and temporal density simultaneously. Similarly, \textit{SCC}, \textit{SABC}, and \textit{Top-r} may achieve acceptable \textit{TD} values on some datasets, but their \textit{TC} and \textit{TWC} remain unstable because the static view only captures aggregated density. The temporal bipartite baselines \textit{PCSearch}, \textit{RCSearch}, and \textit{TABC} also fail to achieve consistently strong performance across all three metrics. Overall, these results show that existing models cannot optimize our objective well, whereas our method achieves a better balance between temporal density and separability. 

\begin{remark}
Optimizing \textit{TWC}, \textit{TD}, and \textit{TC} simultaneously is inherently difficult, since they capture different aspects of community quality. In particular, \textit{TD} emphasizes internal density, whereas \textit{TC} focuses on external separability, and the two are often in conflict. Thus, it is natural that different methods favor different metrics. Since \textit{TWC} is the objective explicitly optimized in our model, our advantage on this metric is expected. As shown in Table~\ref{tab:effectiveness}, no single method consistently dominates all three metrics, while our method achieves the best overall balance among them, which is also consistent with similar observations in~\cite{LinQTCS24}.
\end{remark}

\begin{figure*}[t]
    \centering
    \setlength{\abovecaptionskip}{2pt}
    \setlength{\belowcaptionskip}{0pt}

    \subfigure[QTCS ($U$)]{\includegraphics[width=0.245\linewidth]{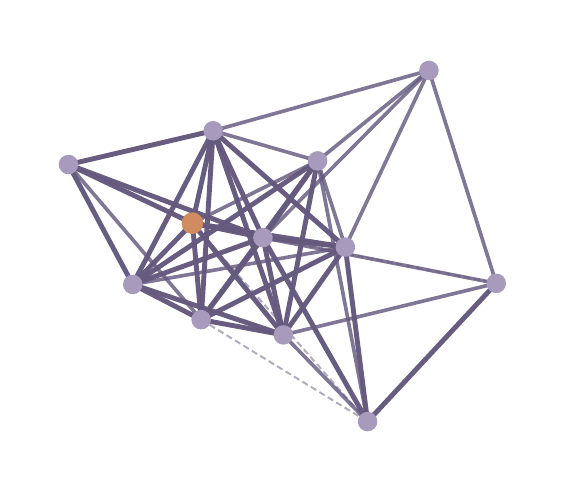}}
    \hspace{-0.8em}
    \subfigure[PCSearch ($U$)]{\includegraphics[width=0.245\linewidth]{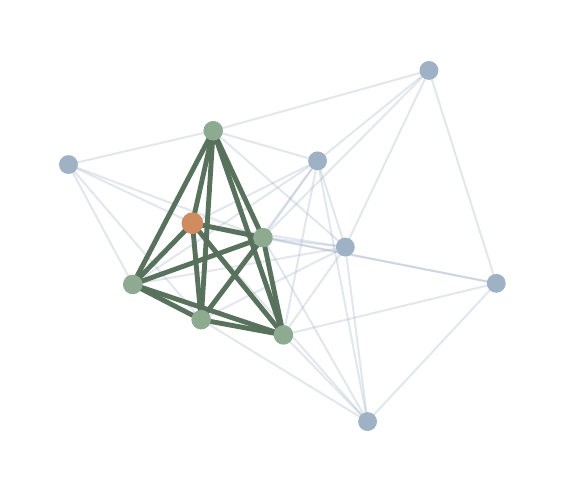}}
    \hspace{-0.8em}
    \subfigure[RCSearch ($U$)]{\includegraphics[width=0.245\linewidth]{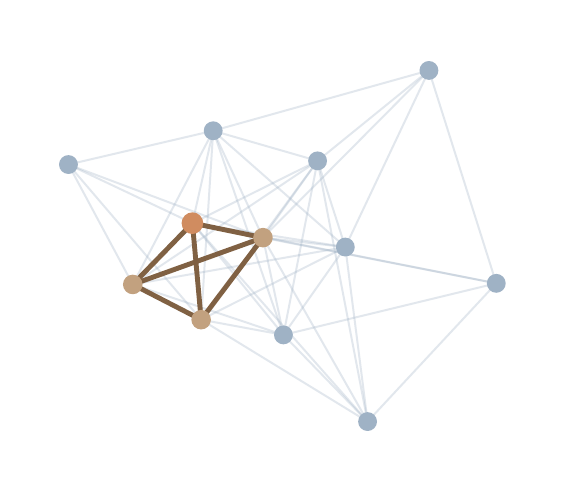}}
    \hspace{-0.8em}
    \subfigure[WCCS ($U$)]{\includegraphics[width=0.245\linewidth]{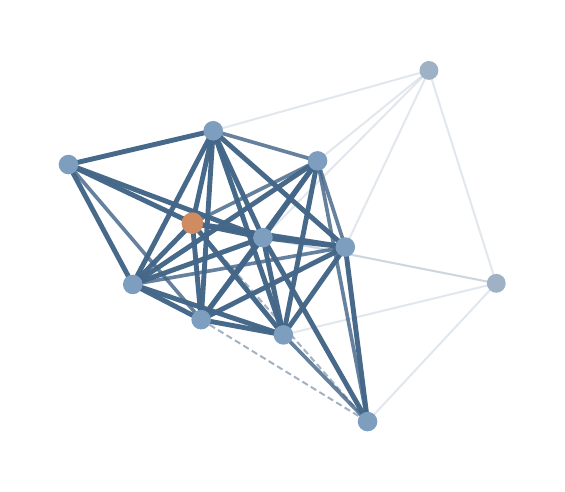}}

    \vspace{-0.9em}

    \subfigure[QTCS ($V$)]{\includegraphics[width=0.245\linewidth]{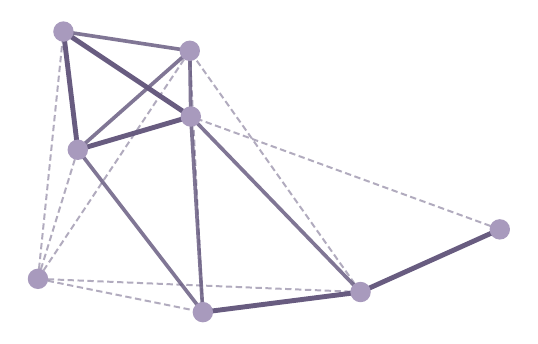}}
    \hspace{-0.8em}
    \subfigure[PCSearch ($V$)]{\includegraphics[width=0.245\linewidth]{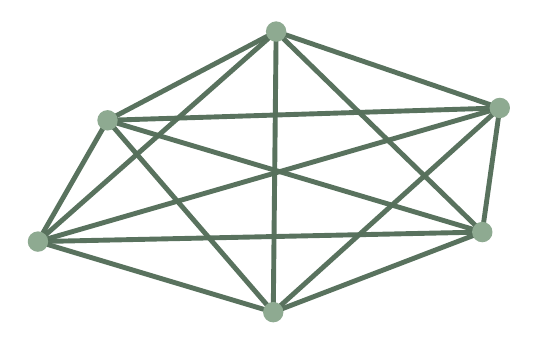}}
    \hspace{-0.8em}
    \subfigure[RCSearch ($V$)]{\includegraphics[width=0.245\linewidth]{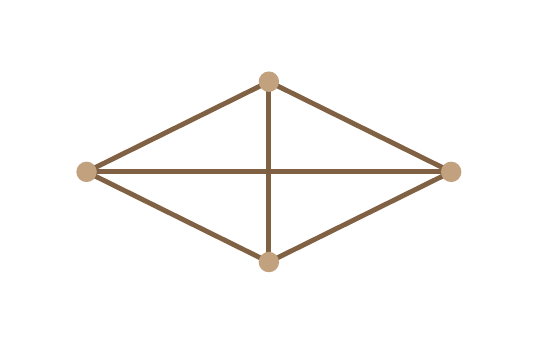}}
    \hspace{-0.8em}
    \subfigure[WCCS ($V$)]{\includegraphics[width=0.245\linewidth]{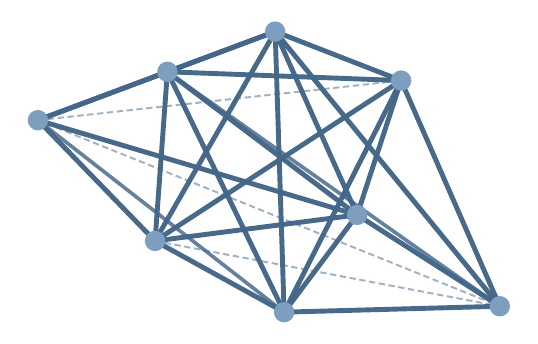}}

    \vspace{-0.4em}
    \caption{Case study on \textit{Ip} over $[1,6]$, where the query node is highlighted in orange. The first row shows the returned $U$-side communities of different methods. In the second row, for \textit{PCSearch} and \textit{RCSearch}, the shown $V$-side nodes are their returned $V$ subsets, while for \textit{QTCS} and \textit{WCCS}, we visualize the corresponding dense subgraphs on the $V$ side induced by wedge-based relations from their returned $U$-side communities. Edge styles encode wedge-based interaction frequency within $[1,6]$: \textit{dashed} for frequency 1, \textit{solid} for frequency 2, and thicker \textit{solid} for frequency at least 3.}
    \label{fig:case_study}
\end{figure*}

\myparagraph{Exp-7: Case studies}
We investigate a query-centered suspicious group inspection task on \textit{Ip}, a customer-product temporal bipartite network where edges denote clicking interactions. Given the query customer marked in orange over the interval $[1,6]$ in Figure~\ref{fig:case_study}, we compare \textit{QTCS}, \textit{PCSearch}, \textit{RCSearch}, and \textit{WCCS}. We omit \textit{MCTS}, \textit{SCC}, and \textit{SABC} because their returned communities are too large to visualize clearly, and exclude \textit{Top-r} and \textit{TABC} because they are not originally designed for query-centered community search. The first row of Figure~\ref{fig:case_study} shows the returned $U$-side communities. In the second row, we directly plot the returned $V$-side subsets of \textit{PCSearch} and \textit{RCSearch}, while for \textit{QTCS} and \textit{WCCS}, we visualize the corresponding dense $V$-side subgraphs induced by wedge-based relations from their returned $U$-side communities.

As shown in Figure~\ref{fig:case_study}, \textit{WCCS} returns the most balanced result. Compared with \textit{QTCS}, its $U$-side community is clearly more compact, while its supporting $V$-side structure still preserves many strong wedge-based interactions. In contrast, \textit{QTCS} returns the largest community in (a), but its corresponding $V$-side subgraph in (e) is much weaker, indicating that many additional vertices are only weakly connected through low-frequency wedge interactions. \textit{PCSearch} and \textit{RCSearch} exhibit the opposite behavior. Their results in (b), (c), (f), and (g) confirm that both methods can identify highly stable cores, but they are also overly conservative and discard many collaborators that still maintain strong yet temporally discrete repeated interactions with the dense center. Therefore, Figure~\ref{fig:case_study} shows that WCCS avoids both the noisy expansion of \textit{QTCS} and the overly strict contraction of \textit{PCSearch} and \textit{RCSearch}, making it better suited for identifying meaningful suspicious groups around the query vertex.

%% file: section/Relatedwork.tex
\section{Related Work}
\label{sec:rel}

\stitle{Community Search over Static Bipartite Graphs.}
Community search over static bipartite graphs has been extensively studied under different structural constraints~\cite{wangICDE21, wang2023significantbp,wangperbc,Chenpercolation23,LiAWCS22,Xvattrib,Zhangbitruss,ZHOUSize23, zhang2024size, LISize24, zhang2150top}. Early studies mainly focus on structural cohesiveness. For example, Wang \etal~\cite{wangICDE21, wang2023significantbp} formalize significant $(\alpha,\beta)$-community search in weighted bipartite graphs, while Wang \etal~\cite{wangperbc} and Chen \etal~\cite{Chenpercolation23} further investigate stricter structures such as personalized maximum biclique search and biclique-based percolation community search. Beyond purely structural settings, Li \etal~\cite{LiAWCS22} and Xv \etal~\cite{Xvattrib} incorporate vertex attributes into bipartite community search. Zhang \etal~\cite{zhang2150top} further study top-$r$ influential community search in bipartite graphs, which aims to retrieve multiple influential communities rather than a single query-centered one. More recently, studies such as~\cite{ZHOUSize23, zhang2024size, LISize24} further introduce size constraints and develop peeling-based or expansion-oriented methods to refine the returned community. However, all these approaches are designed for static bipartite graphs and ignore temporal information, which limits their applicability in temporal settings (Section~\ref{exp}).

\stitle{Community Search over Temporal Graphs.}
In temporal unipartite graphs, Li \etal~\cite{LiCSS21} study community continuity by identifying a $(\theta,\tau)$-continual $k$-core that remains densely connected over the long term, and develop both exact and approximate algorithms for this NP-hard problem. Li \etal~\cite{LIactive} further investigate the most active community search by maximizing time-decayed activity, while Zhang \etal~\cite{ZhangEngage22} focus on communities with high cumulative engagement over time. To mitigate the free-rider issue, Lin \etal~\cite{LinQTCS24} formulate query-centered temporal community search and propose a time-constrained Personalized PageRank together with a $\beta$-temporal proximity core model. Zhong \etal~\cite{Zhong25kcore} further study temporal $k$-core based community search, while Yang \etal~\cite{Yangtruss} and Du \etal~\cite{DuMTZ23} extend the setting to truss-based and attribute-constrained temporal communities, respectively. However, these methods cannot be directly applied to our problem because they ignore the heterogeneous interactions inherent in bipartite graphs.

For temporal bipartite graphs, Li \etal~\cite{li2023persistent} devise a $(\theta,\tau)$-persistent $(\alpha,\beta)$-core model and develop the branch--reduce--bound style PCSearch algorithm, which measures persistence over continuous time intervals and is therefore unsuitable for snapshot-based frequency verification. Li \etal~\cite{LiMoreliable} further propose a reliable $(\alpha,\beta)$-community model that jointly considers time span, size, and degree constraints. Tian \etal~\cite{tabc24} study temporal core queries over bipartite graphs and develop an efficient index for retrieving temporally cohesive cores. Nevertheless, these methods mainly focus on interval-based persistence or temporal core retrieval. Their scoring schemes emphasize structural cohesion and temporal validity, while neglecting external sparsity. Moreover, they still rely on edge-based cohesive models and cannot capture the higher-order wedge interactions considered in our work, which leads to suboptimal performance (Section~\ref{exp}).

%% file: section/Conclusion.tex

\section{Conclusion}
\label{sec:conc}

We proposed a novel frequency-aware community search model over temporal bipartite graphs, Wedge Conductance Community Search (WCCS), which explicitly characterizes both internal density and external sparsity, which are two crucial factors largely overlooked in existing methods. By introducing the higher-order $(\alpha,\beta,\tau)$-wedge core and the temporal wedge conductance metric, our model effectively overcomes the limitations of prior edge-centric and interval-based approaches. We further designed efficient online algorithms with conductance-guided search and dynamic programming techniques, as well as an offline compressed index to ensure scalability. Extensive experiments on  seven datasets show that our solutions significantly outperform eight competitors
in terms of efficiency, scalability, and effectiveness.

%% file: section/Appendix.tex
\appendix
\subsection{Missing Proofs}

\stitle{Proof of Theorem \ref{th:NP-Hard}.}
We prove the NP-hardness of WCCS by a polynomial-time Turing reduction from MCC~\cite{fiedler1973algebraic,DBLP:conf/stoc/SpielmanT04,DBLP:conf/focs/AndersenCL06,DBLP:conf/www/LeskovecLM10,DBLP:conf/aaai/LinLJ23}. Given an undirected graph $H=(V_H,E_H)$, we construct a temporal bipartite graph $\mathcal{G}=(U,V,\mathcal{E})$ with a single timestamp $t_0$ by letting $U=V_H$ and, for each edge $e=\{x,y\}\in E_H$, creating two vertices $e^{(1)},e^{(2)}\in V$ both adjacent to $x$ and $y$, with all incidences assigned to $t_0$. Let $\alpha=\beta=\tau=1$. MCC admits a connected optimum because if $H[S]$ is disconnected, then $\phi_H(S)$ is a weighted average of the conductance values of the connected components of $H[S]$, so some component has conductance no larger than $\phi_H(S)$. Now let $S\subseteq U$ be any connected set with $|S|\ge 2$, and define $V_S=\{e^{(1)},e^{(2)}\in V\mid e\in E_H(S)\}$. Then $S$ is a feasible $(1,1,1)$-wedge core in $\mathcal{G}$ because every $u\in S$ has at least one wedge neighbor in $S$, every $v\in V_S$ has at least one wedge neighbor via its duplicate, and the induced bipartite subgraph on $(S,V_S)$ is connected. Moreover, each edge of $H$ contributes exactly two duplicated support vertices, thus $\mathrm{vol}_w(S,t_0)=2\,\mathrm{vol}_H(S)$, $\mathrm{cut}_w(S,t_0)=2\,\mathrm{cut}_H(S)$, and $\mathbf{W}_{t_0}-\mathrm{vol}_w(S,t_0)=2\,\mathrm{vol}_H(V_H\setminus S)$. Hence $\phi_w(S,t_0)=\phi_H(S)$, and since $\mathcal{T}=\{t_0\}$ and $\tau=1$, we further have $\Phi(S)=\phi_H(S)$. Suppose WCCS can be solved in polynomial time. Then, for every query vertex $q\in U$, we run WCCS on $(\mathcal{G},q,\alpha,\beta,\tau)$ and return the feasible solution with minimum objective value over all runs. Let $S^*$ be a connected optimum of MCC. For every $q\in S^*$, $S^*$ is a feasible WCCS solution containing $q$ and satisfies $\Phi(S^*)=\phi_H(S^*)$. Therefore, one of these polynomially many runs returns an optimum of MCC, which is impossible unless $\mathrm{P}=\mathrm{NP}$. Thus, WCCS is NP-hard. The same reduction transfers any polynomial-time $\gamma$-approximation for WCCS to MCC. Since MCC admits no polynomial-time constant-factor approximation unless $\mathrm{P}=\mathrm{NP}$~\cite{SimaS06}, neither does WCCS.

\stitle{Proof of Lemma \ref{le:qa-filter}.}
Since $C$ is a valid $(\alpha,\beta,\tau)$-wedge core, by Definition \ref{abc}, there exists a valid timestamp set $\mathcal{T}_C\subseteq \mathcal{T}_Q$ with $|\mathcal{T}_C|\ge \tau$ such that, for every $t\in \mathcal{T}_C$, the community $C$ forms an $(\alpha,\beta)$-wedge core containing $q$ in $G_t$. Hence, for every $u\in C$ and every $t\in \mathcal{T}_C$, we have $u\in \mathcal{W}_{\alpha,\beta}(G_t,q)$. Therefore, $u$ appears in $\mathcal{W}_{\alpha,\beta}(G_t,q)$ in at least $\tau$ snapshots within $\mathcal{T}_Q$.

\stitle{Proof of Lemma \ref{le:convex}.} If $\mathrm{slope}_C(b,c)\le \mathrm{slope}_C(a,b)$, then $(b,CWC[C][b])$ lies on or above the line segment joining $(a,CWC[C][a])$ and $(c,CWC[C][c])$. For any future endpoint $t>c$, if $\mathrm{slope}_C(b,t)\le \mathrm{slope}_C(b,c)$, then $\mathrm{slope}_C(b,t)\le \mathrm{slope}_C(a,b)$, which implies $\mathrm{slope}_C(a,t)\ge \mathrm{slope}_C(b,t)$. Otherwise, if $\mathrm{slope}_C(b,t)>\mathrm{slope}_C(b,c)$, then $\mathrm{slope}_C(c,t)>\mathrm{slope}_C(b,t)$. In either case, index $b$ cannot be optimal for any future endpoint and is therefore redundant.

\stitle{Proof of Lemma \ref{le:head}.} The condition $\mathrm{slope}_C(a,j)\le \mathrm{slope}_C(b,j)$ implies that the optimal starting index for endpoint $j$ is at least $b$. Since the optimal point on the lower convex hull is monotone with respect to the endpoint, the optimal starting index is non-decreasing as $j$ increases. Therefore, for any later endpoint $j'>j$, index $a$ can never become optimal again and is permanently redundant.

\stitle{Proof of Lemma \ref{mono-support}.}
For any support vertex adjacent to $x$, its wedge degree with respect to $C'$ at timestamp $t$ may increase because $x$ introduces additional wedges; for vertices not adjacent to $x$, the wedge degree remains unchanged. Therefore, every vertex in $V_t(C)$ that already satisfies the $\beta$-constraint remains valid in $V_t(C')$, while previously invalid support vertices may become valid after inserting $x$. Hence $V_t(C)\subseteq V_t(C')$. Since the support set can only expand, the wedge degree of each $u\in C$ is non-decreasing.

\stitle{Proof of Lemma \ref{eq:cut-update}.} 
All wedges incident to $x$ are partitioned into two groups: those whose other endpoint lies in $C$, and those whose other endpoint lies in $U\setminus C'$. Hence,
$\mathrm{cut}_w(x,C,t)+\mathrm{cut}_w(x,U\setminus C',t)=d_w(x,G_t).$
After inserting $x$ into $C$, wedges between $x$ and $C$ become internal, while wedges between $x$ and $U\setminus C'$ contribute to the new cut. Therefore,
$\mathrm{cut}_w(C',t) = \mathrm{cut}_w(C,t)-\mathrm{cut}_w(x,C,t)+\mathrm{cut}_w(x,U\setminus C',t) = \mathrm{cut}_w(C,t)+d_w(x,G_t)-2\,\mathrm{cut}_w(x,C,t).$

\stitle{Proof of Lemma \ref{lem:subtree-lb}.} 
For any feasible descendant $C'\supseteq C$, monotonicity gives $\mathcal{T}_{C'}\subseteq \mathcal{CT}_{C'}\subseteq \mathcal{CT}_C$. Moreover, for any $x\in \mathcal{B}(C)$, we have $|\mathcal{CT}_{C'}\cap \mathcal{CT}(x)|\le |\mathcal{CT}_C\cap \mathcal{CT}(x)|<\tau$, implying $x\notin C'$. Together with $C'\cap \mathcal{F}=\emptyset$, this yields $C'\cap \mathcal{I}(C)=\emptyset$. Therefore, every wedge counted by $\underline{\mathrm{cut}}_w(C,t)$ must remain a cut wedge for any feasible descendant $C'$, and hence $\mathrm{cut}_w(C',t)\ge \underline{\mathrm{cut}}_w(C,t)$. Since $\min(\mathrm{vol}_w(C',t),\,\mathbf{W}_t-\mathrm{vol}_w(C',t))\le \mathbf{W}_t/2$, it follows that $\phi_w(C',t)\ge \underline{\phi}(C,t)$ for every $t\in \mathcal{T}_{C'}$. Let $\mathcal{T}_{C'}=\{t_{(1)},\dots,t_{(L)}\}$ with $L\ge\tau$. By Definition \ref{twc}, we have $\Phi(C') \ge \sum_{k=1}^{L}\phi_w(C',t_{(k)}) /L \ge \sum_{k=1}^{L}\underline{\phi}(C,t_{(k)})/L \ge \mathrm{LB}(C).$
Thus, if $\mathrm{LB}(C)\ge \Phi^*$, no feasible descendant of $C$ can improve the current best solution, and the whole branch can be pruned safely.

\stitle{Proof of Theorem \ref{the:timecomplexity}.}
The total running time is the aggregate cost of processing all visited states in the priority queue. For each state $C \in \mathcal{S}$, the algorithm iterates through its candidate set. For each candidate $u$, it verifies the wedge constraints across the $\mathcal{CT}$ intersection bounded by $|\mathcal{T}|$. The \textit{CheckWCore} procedure at a specific snapshot $t$ requires enumerating the wedges incident to $u$, which takes $O(d_{max}(U)\cdot d_{max}(V))$ time. Therefore, the total complexity is $T_{total} = \sum_{C \in \mathcal{S}} \sum_{u \in Cand(C)} \sum_{t \in \mathcal{CT}_{C'}} O\!\left(d_{max}(U) \cdot d_{max}(V)\right)
= O\left(|\mathcal{S}| \cdot n \cdot |\mathcal{T}| \cdot d_{max}(U) \cdot d_{max}(V)\right)$.

\stitle{Proof of Theorem \ref{the:WCI}.}
Consider a snapshot $G_t$. The common-neighbor entries in $\mathrm{coU}$ and $\mathrm{coV}$ are initialized by enumerating all two-hop pairs induced by common neighbors in Algorithm \ref{alg:Skyline}. The total number of such pairs is $\sum_{v\in V}\binom{d_t(v)}{2}+\sum_{u\in U}\binom{d_t(u)}{2}$. During peeling, whenever a vertex is removed, only the entries supported by this vertex are updated, and each update decreases the corresponding counter by one. Hence, every unit counted in the above two-hop enumeration is processed at most once throughout the whole peeling procedure. Therefore, Algorithm \ref{alg:Skyline} takes $O\!\left(\sum_{v\in V}\binom{d_t(v)}{2}+\sum_{u\in U}\binom{d_t(u)}{2}\right)$ time on $G_t$. Using $\sum_{v\in V}\binom{d_t(v)}{2}+\sum_{u\in U}\binom{d_t(u)}{2}=\frac{1}{2}\left(\sum_{u\in U} d_w(u,G_t)+\sum_{v\in V} d_w(v,G_t)\right)$ and $d_w(x,G_t)\le \Delta\, d_t(x)$ for any $x\in U\cup V$ in $G_t$, summing over all snapshots gives $\sum_{t\in\mathcal{T}}O(\Delta |E_t|)=O(\Delta m)$. The materialization phase scans each recorded skyline membership and each resulting accumulator entry once. Therefore, its time complexity is linear in the total number of skyline memberships generated over all snapshots, namely $O\!\left(\sum_{t\in\mathcal{T}}\left(\sum_{u\in U}|\mathcal{K}_t(u)|+\sum_{v\in V}|\mathcal{K}_t(v)|\right)\right)$. Since all these skyline memberships are generated during the snapshot-wise peeling process, this quantity is asymptotically bounded by the total skyline computation cost, and thus the materialization phase also takes $O(\Delta m)$ time. Combining the two phases yields the overall complexity $O(\Delta m)$.

\subsection{Additional Experiments}

\begin{figure*}[t]
    \centering
    \subfigure[Ip (varying $\alpha$)]{\includegraphics[width=0.24\linewidth]{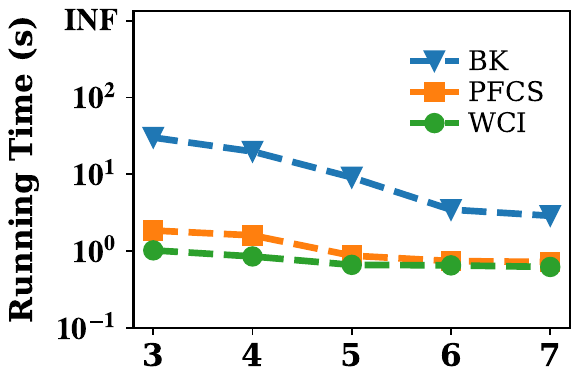}}
    \subfigure[LK (varying $\alpha$)]{\includegraphics[width=0.24\linewidth]{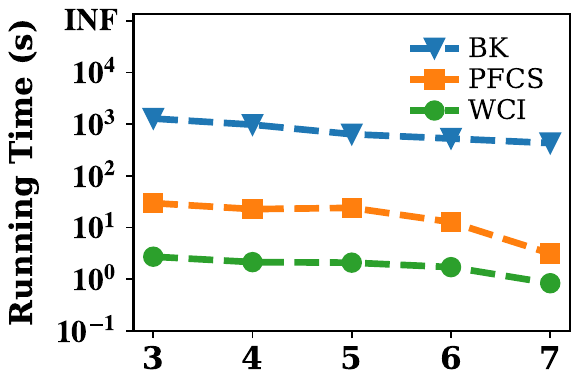}}
    \subfigure[Bti  (varying $\alpha$)]{\includegraphics[width=0.24\linewidth]{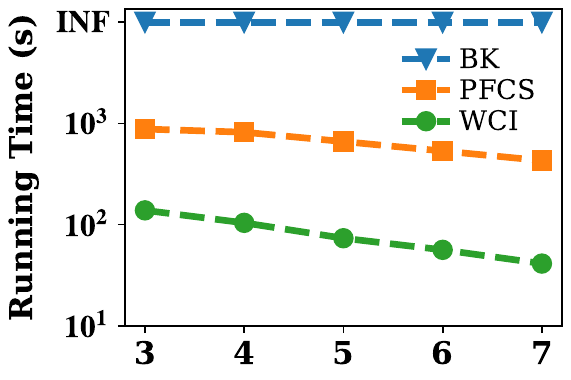}}
    \subfigure[ar (varying $\alpha$)]{\includegraphics[width=0.24\linewidth]{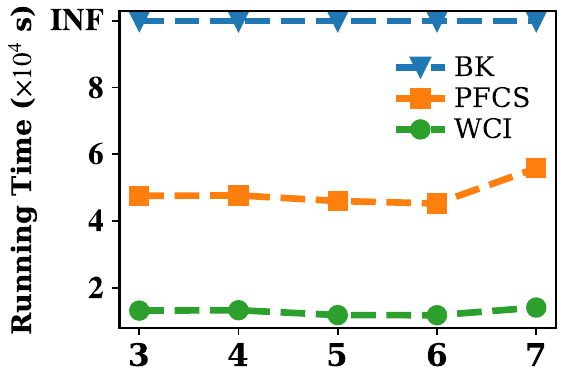}}
    \subfigure[Ip (varying $\beta$)]{\includegraphics[width=0.24\linewidth]{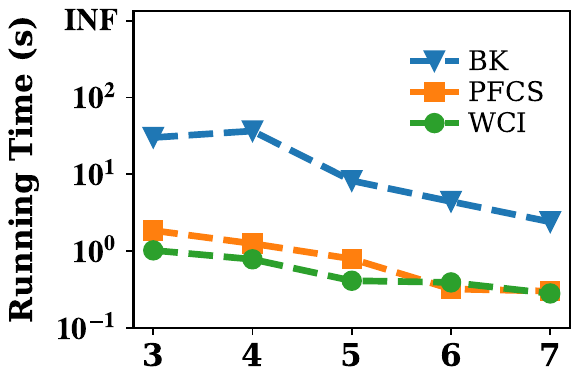}}
    \subfigure[LK (varying $\beta$)]{\includegraphics[width=0.24\linewidth]{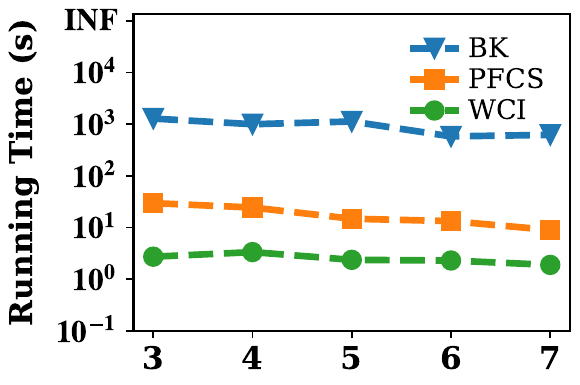}}
    \subfigure[Bti (varying $\beta$)]{\includegraphics[width=0.24\linewidth]{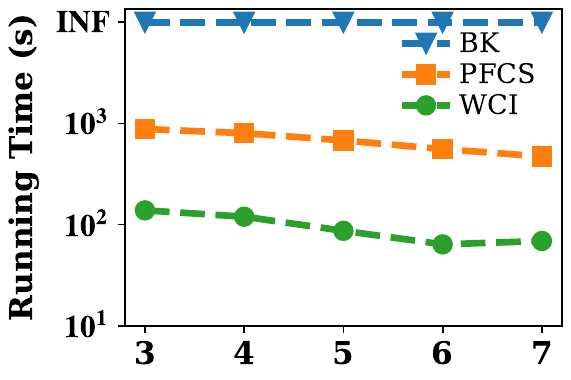}}
    \subfigure[ar (varying $\beta$)]{\includegraphics[width=0.24\linewidth]{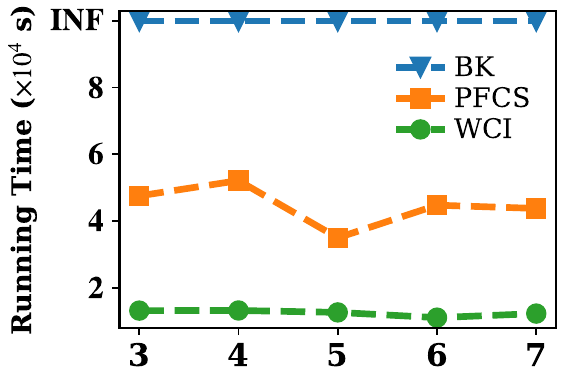}}
    \subfigure[Ip (varying $\tau$)]{\includegraphics[width=0.24\linewidth]{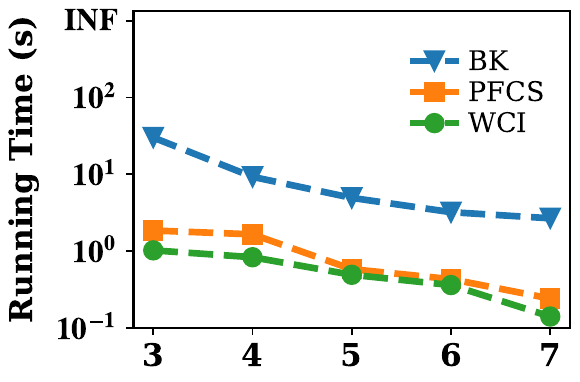}}
    \subfigure[LK (varying $\tau$)]{\includegraphics[width=0.24\linewidth]{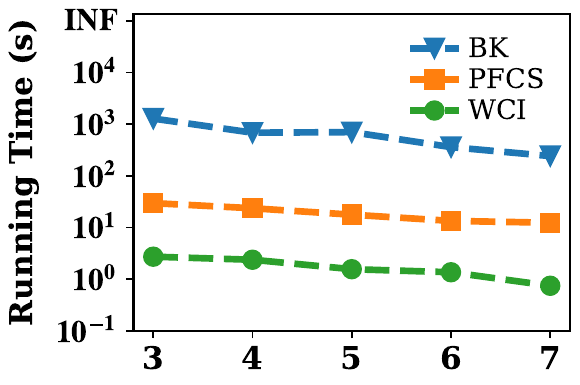}}
    \subfigure[Bti (varying $\tau$)]{\includegraphics[width=0.24\linewidth]{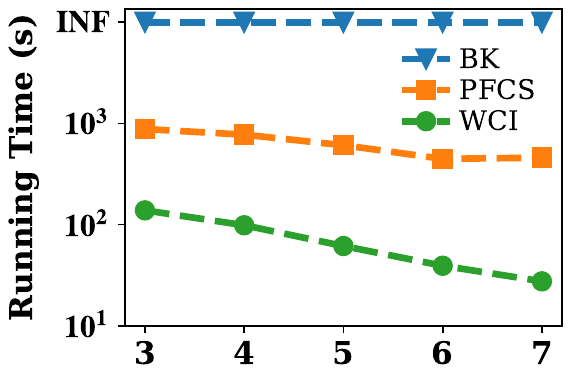}}
    \subfigure[ar (varying $\tau$)]{\includegraphics[width=0.23\linewidth]{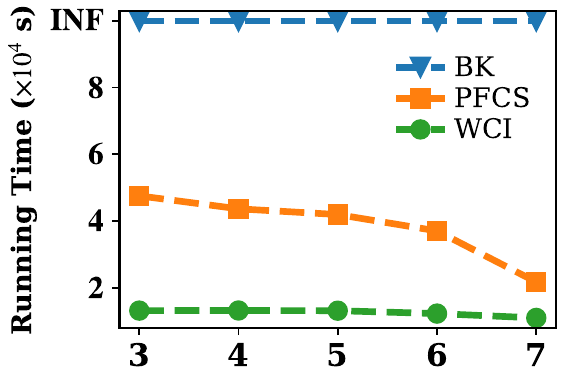}}
    \caption{Running time of different WCCS algorithms with varying parameters.}
    \label{fig:exp2}
\end{figure*}

\begin{table*}[t]
\centering
\renewcommand{\arraystretch}{1.12}
\caption{Community quality of WCCS under varying $\alpha$, $\beta$, and $\tau$. Each entry is reported as TWC / TC / TD, where smaller TWC and TC indicate better quality, while larger TD indicates better quality.}
\label{tab:param_quality}
\begin{adjustbox}{width=0.95\textwidth,center}
\begin{tabular}{cc!{\vrule width 0.8pt}ccccccc}
\toprule
Parameter & Value & \textit{Ip} & \textit{diq} & \textit{vec} & \textit{LK} & \textit{Wut} & \textit{Bti} & \textit{ar} \\
\midrule

\multirow{5}{*}{$\alpha$}
& 3 & 0.09 / 0.32 / 0.59 & 0.22 / 0.58 / 0.57 & 0.35 / 0.53 / 0.61 & 0.07 / 0.33 / 0.41 & 0.46 / 0.59 / 0.55 & 0.49 / 0.52 / 0.66 & 0.61 / 0.57 / 0.58 \\
& 4 & 0.10 / 0.31 / 0.48 & 0.21 / 0.54 / 0.59 & 0.34 / 0.52 / 0.62 & 0.08 / 0.35 / 0.43 & 0.44 / 0.57 / 0.57 & 0.37 / 0.42 / 0.71 & 0.59 / 0.55 / 0.67 \\
& 5 & 0.22 / 0.44 / 0.63 & 0.19 / 0.50 / 0.62 & 0.31 / 0.48 / 0.65 & 0.09 / 0.34 / 0.44 & 0.41 / 0.54 / 0.59 & 0.44 / 0.48 / 0.68 & 0.53 / 0.49 / 0.62 \\
& 6 & 0.18 / 0.27 / 0.46 & 0.20 / 0.52 / 0.61 & 0.32 / 0.49 / 0.64 & 0.07 / 0.31 / 0.56 & 0.42 / 0.56 / 0.60 & 0.45 / 0.49 / 0.69 & 0.56 / 0.52 / 0.70 \\
& 7 & 0.16 / 0.32 / 0.55 & 0.17 / 0.45 / 0.66 & 0.28 / 0.44 / 0.68 & 0.08 / 0.29 / 0.47 & 0.38 / 0.49 / 0.62 & 0.41 / 0.44 / 0.72 & 0.63 / 0.61 / 0.55 \\

\midrule

\multirow{5}{*}{$\beta$}
& 3 & 0.09 / 0.32 / 0.59 & 0.22 / 0.58 / 0.57 & 0.35 / 0.53 / 0.61 & 0.07 / 0.33 / 0.41 & 0.46 / 0.59 / 0.55 & 0.49 / 0.52 / 0.66 & 0.61 / 0.57 / 0.58 \\
& 4 & 0.10 / 0.29 / 0.41 & 0.34 / 0.48 / 0.67 & 0.41 / 0.49 / 0.55 & 0.07 / 0.34 / 0.45 & 0.44 / 0.57 / 0.58 & 0.38 / 0.49 / 0.62 & 0.60 / 0.56 / 0.59 \\
& 5 & 0.10 / 0.30 / 0.41 & 0.31 / 0.52 / 0.61 & 0.32 / 0.50 / 0.64 & 0.10 / 0.29 / 0.43 & 0.41 / 0.54 / 0.59 & 0.45 / 0.48 / 0.69 & 0.56 / 0.52 / 0.62 \\
& 6 & 0.09 / 0.26 / 0.43 & 0.28 / 0.56 / 0.63 & 0.30 / 0.47 / 0.66 & 0.09 / 0.31 / 0.50 & 0.38 / 0.49 / 0.64 & 0.41 / 0.45 / 0.72 & 0.53 / 0.49 / 0.65 \\
& 7 & 0.10 / 0.31 / 0.48 & 0.37 / 0.54 / 0.59 & 0.34 / 0.42 / 0.59 & 0.08 / 0.35 / 0.43 & 0.43 / 0.55 / 0.58 & 0.46 / 0.48 / 0.69 & 0.58 / 0.53 / 0.61 \\

\midrule

\multirow{5}{*}{$\tau$}
& 3 & 0.09 / 0.32 / 0.59 & 0.22 / 0.58 / 0.57 & 0.35 / 0.53 / 0.61 & 0.07 / 0.33 / 0.41 & 0.46 / 0.59 / 0.55 & 0.49 / 0.52 / 0.66 & 0.61 / 0.57 / 0.58 \\
& 4 & 0.10 / 0.31 / 0.48 & 0.21 / 0.47 / 0.53 & 0.36 / 0.54 / 0.60 & 0.15 / 0.42 / 0.51 & 0.43 / 0.57 / 0.57 & 0.47 / 0.51 / 0.68 & 0.60 / 0.56 / 0.59 \\
& 5 & 0.28 / 0.49 / 0.71 & 0.31 / 0.63 / 0.50 & 0.33 / 0.51 / 0.63 & 0.23 / 0.43 / 0.60 & 0.39 / 0.51 / 0.63 & 0.52 / 0.55 / 0.65 & 0.66 / 0.60 / 0.56 \\
& 6 & 0.27 / 0.42 / 0.73 & 0.22 / 0.45 / 0.61 & 0.27 / 0.56 / 0.59 & 0.19 / 0.39 / 0.56 & 0.42 / 0.54 / 0.58 & 0.45 / 0.49 / 0.70 & 0.57 / 0.52 / 0.61 \\
& 7 & 0.23 / 0.45 / 0.69 & 0.41 / 0.52 / 0.63 & 0.34 / 0.52 / 0.61 & 0.21 / 0.41 / 0.54 & 0.31 / 0.40 / 0.64 & 0.54 / 0.54 / 0.66 & 0.65 / 0.57 / 0.59 \\

\bottomrule
\end{tabular}
\end{adjustbox}
\end{table*}

\myparagraph{Running time by varying parameters}
To evaluate the effect of parameter settings, we select four representative datasets with different scales, i.e., \textit{Ip}, \textit{LK}, \textit{Bti}, and \textit{ar}. Figures \ref{fig:exp2}(a) to \ref{fig:exp2}(l) report the response time on these four datasets by varying $\alpha$, $\beta$, and $\tau$, respectively. As can be seen: (1) \textit{WCI} consistently achieves the highest efficiency under all settings, \textit{PFCS} always ranks second, and \textit{BK} is the slowest whenever it can finish. This is particularly evident on \textit{Bti} and \textit{ar}, while both \textit{PFCS} and \textit{WCI} remain efficient. (2) The curves of \textit{PFCS} and \textit{WCI} are generally smoother than those of \textit{BK}, especially on \textit{Ip}. For example, when varying $\alpha$ or $\tau$, the running time of \textit{BK} drops much more sharply, whereas \textit{PFCS} and \textit{WCI} change more gradually. (3) The parameter $\tau$ has the strongest impact on running time, as shown in the third row of Figure \ref{fig:exp2}, especially for \textit{Ip} and \textit{ar}, where the slopes are steeper. This is because $\tau$ directly constrains temporal validity and shrinks the feasible search space more aggressively than $\alpha$ and $\beta$. Despite this stronger effect, the curve of \textit{WCI} remains slightly smoother than that of \textit{PFCS} for most datasets and significantly more stable than that of \textit{BK}. In summary, \textit{BK} is more directly affected by the change of the feasible search space, while our methods control the query cost more effectively through pruning and incremental maintenance.

\myparagraph{Evaluating community quality of \textit{WCCS} under varying parameters}
Table \ref{tab:param_quality} reports the community quality of \textit{WCCS} under different settings of $\alpha$, $\beta$, and $\tau$. Overall, the three metrics vary smoothly across most datasets, indicating that \textit{WCCS} is relatively insensitive to moderate parameter changes and can consistently preserve favorable community quality. As can be seen, (1) as $\alpha$ varies, the quality changes are generally limited, e.g., \textit{TWC} decreases from 0.35 to 0.28, \textit{TC} from 0.53 to 0.44, and \textit{TD} increases from 0.61 to 0.68 on \textit{vec}, while the three metrics on \textit{Wut} also remain within a narrow interval. This suggests that strengthening the $U$-side structural requirement usually leads to slightly denser communities without fundamentally altering the result quality. (2) The effect of $\beta$ is similarly stable, with most datasets showing only mild fluctuations. For instance, on \textit{Ip}, \textit{TWC} remains between 0.09 and 0.10, \textit{TC} between 0.26 and 0.32, while \textit{TD} changes only from 0.41 to 0.48. (3) Among the three parameters, $\tau$ has the most noticeable influence. This is particularly visible on \textit{Ip} and \textit{LK}, where both \textit{TWC} and \textit{TC} increase more evidently as $\tau$ becomes stricter, while \textit{TD} still stays in a relatively stable range. Nevertheless, the overall variations remain moderate rather than disruptive. In summary, these results confirm that \textit{WCCS} maintains stable community quality under different structural and temporal constraints.

%% file: main.bib
@article{DBLP:journals/tii/ChenLLJ26,
  author  = {Zhixuan Chen and Xinyu Liu and Longlong Lin and Tao Jia},
  title   = {Effective and Efficient Temporal Graph Neural Networks via Polynomial Spectral Sparsification},
  journal = {IEEE Transactions on Industrial Informatics},
  volume  = {22},
  number  = {8},
  pages   = {7417--7428},
  year    = {2026},
  doi     = {10.1109/TII.2026.3686195}
}

@inproceedings{DBLP:conf/icde/LengZQLL25,
  author    = {Xiaoyu Leng and Guang Zeng and Hongchao Qin and Longlong Lin and Rong-Hua Li},
  title     = {On Temporal-Constraint Subgraph Matching},
  booktitle = {Proceedings of the 41st IEEE International Conference on Data Engineering},
  pages     = {2493--2506},
  publisher = {IEEE},
  year      = {2025},
  doi       = {10.1109/ICDE65448.2025.00188}
}

@article{DBLP:journals/tsmc/LinYLWLJ22,
  author  = {Longlong Lin and Pingpeng Yuan and Rong-Hua Li and Jifei Wang and Ling Liu and Hai Jin},
  title   = {Mining Stable Quasi-Cliques on Temporal Networks},
  journal = {IEEE Transactions on Systems, Man, and Cybernetics: Systems},
  volume  = {52},
  number  = {6},
  pages   = {3731--3745},
  year    = {2022},
  doi     = {10.1109/TSMC.2021.3071721}
}

@article{DBLP:journals/tbd/LinYLJ22,
  author  = {Longlong Lin and Pingpeng Yuan and Rong-Hua Li and Hai Jin},
  title   = {Mining Diversified Top-r Lasting Cohesive Subgraphs on Temporal Networks},
  journal = {IEEE Transactions on Big Data},
  volume  = {8},
  number  = {6},
  pages   = {1537--1549},
  year    = {2022},
  doi     = {10.1109/TBDATA.2021.3058294}
}

@inproceedings{DBLP:conf/ppopp/Hou0H00YL00L026,
  author    = {Yinbo Hou and Hao Qi and Ligang He and Jin Zhao and Yu Zhang and Hui Yu and Longlong Lin and Lin Gu and Wenbin Jiang and Xiaofei Liao and Hai Jin},
  title     = {DTMiner: A Data-Centric System for Efficient Temporal Motif Mining},
  booktitle = {Proceedings of the 31st ACM SIGPLAN Annual Symposium on Principles and Practice of Parallel Programming},
  pages     = {591--604},
  publisher = {ACM},
  year      = {2026},
  doi       = {10.1145/3774934.3786416}
}

@inproceedings{DBLP:conf/asplos/0003WH0DHWY0LYL25,
  author    = {Jin Zhao and Qian Wang and Ligang He and Yu Zhang and Sheng Di and Bingsheng He and Xinlei Wang and Hui Yu and Hao Qi and Longlong Lin and Linchen Yu and Xiaofei Liao and Hai Jin},
  title     = {TempGraph: An Efficient Chain-Driven Temporal Graph Computing Framework on the GPU},
  booktitle = {Proceedings of the 30th ACM International Conference on Architectural Support for Programming Languages and Operating Systems},
  pages     = {230--246},
  publisher = {ACM},
  year      = {2025},
  doi       = {10.1145/3676642.3736116}
}

@IEEEtranBSTCTL{IEEEexample:BSTcontrol,
  CTLdash_repeated_names = "no"
}

@inproceedings{alphabetacoreCIKM,
  author    = {Ding, Danhao and Li, Hui and Huang, Zhipeng and Mamoulis, Nikos},
  title     = {Efficient Fault-Tolerant Group Recommendation Using alpha-beta-core},
  booktitle = {CIKM},
  year      = {2017}
}

@article{DBLP:journals/tcyb/XueNLWL24,
  author       = {Jingjing Xue and
                  Feiping Nie and
                  Chaodie Liu and
                  Rong Wang and
                  Xuelong Li},
  title        = {Co-Clustering by Directly Solving Bipartite Spectral Graph Partitioning},
  journal      = {{IEEE} Trans. Cybern.},
  volume       = {54},
  number       = {12},
  pages        = {7590--7601},
  year         = {2024},
}

@inproceedings{aslay2018mining,
  title     = {Mining Frequent Patterns in Evolving Graphs},
  author    = {Aslay, Cigdem and Nasir, Muhammad Anis Uddin and De Francisci Morales, Gianmarco and Gionis, Aristides},
  booktitle = {CIKM},
  year      = {2018}
}

@article{bron1973algorithm,
  author  = {Bron, Coenraad and Kerbosch, Joep},
  title   = {Finding All Cliques of an Undirected Graph (Algorithm 457)},
  journal = {Communications of the ACM},
  volume  = {16},
  number  = {9},
  year    = {1973}
}

@article{cai2023efficient,
  title   = {Efficient Temporal Butterfly Counting and Enumeration on Temporal Bipartite Graphs},
  author  = {Cai, Xinwei and Ke, Xiangyu and Wang, Kai and Chen, Lu and Zhang, Tianming and Liu, Qing and Gao, Yunjun},
  journal = {Proceedings of the VLDB Endowment},
  year    = {2024}
}

@article{ChenBreahability,
  title   = {Efficiently Answering Reachability and Path Queries on Temporal Bipartite Graphs},
  author  = {Chen, Xiaoshuang and Wang, Kai and Lin, Xuemin and Zhang, Wenjie and Qin, Lu and Zhang, Ying},
  journal = {Proceedings of the VLDB Endowment},
  volume  = {14},
  number  = {10},
  year    = {2021}
}

@INPROCEEDINGS{Chenpercolation23,
  author    = {Chen, Zi and Zhao, Yiwei and Yuan, Long and Lin, Xuemin and Wang, Kai},
  title     = {Index-Based Biclique Percolation Communities Search on Bipartite Graphs},
  booktitle = {ICDE},
  year      = {2023},
  pages     = {2699--2712}
}

@inproceedings{DBLP:conf/aaai/LinLJ23,
  author    = {Lin, Longlong and Li, Ronghua and Jia, Tao},
  title     = {Scalable and Effective Conductance-Based Graph Clustering},
  booktitle = {AAAI},
  pages     = {4471--4478},
  year      = {2023}
}

@inproceedings{DBLP:conf/focs/AndersenCL06,
  author    = {Andersen, Reid and Chung, Fan R. K. and Lang, Kevin J.},
  title     = {Local Graph Partitioning Using PageRank Vectors},
  booktitle = {FOCS},
  pages     = {475--486},
  year      = {2006}
}

@inproceedings{DBLP:conf/stoc/SpielmanT04,
  author    = {Spielman, Daniel A. and Teng, Shang-Hua},
  title     = {Nearly-linear Time Algorithms for Graph Partitioning, Graph Sparsification, and Solving Linear Systems},
  booktitle = {STOC},
  year      = {2004}
}

@inproceedings{DBLP:conf/www/LeskovecLM10,
  author    = {Leskovec, Jure and Lang, Kevin J. and Mahoney, Michael W.},
  title     = {Empirical Comparison of Algorithms for Network Community Detection},
  booktitle = {WWW},
  pages     = {631--640}
}

@article{DBLP:journals/pacmmod/KamalB24,
  author  = {Kamal, Raj and Bagchi, Amitabha},
  title   = {A Lov{\'a}sz-Simonovits Theorem for Hypergraphs with Application to Local Clustering},
  journal = {Proceedings of the ACM on Management of Data},
  volume  = {2},
  number  = {4},
  pages   = {190:1--190:27},
  year    = {2024}
}

@article{fiedler1973algebraic,
  title   = {Algebraic Connectivity of Graphs},
  author  = {Fiedler, Miroslav},
  journal = {Czechoslovak Mathematical Journal},
  volume  = {23},
  number  = {2},
  pages   = {298--305},
  year    = {1973}
}

@inproceedings{Ga2018span,
  author    = {Galimberti, Edoardo and Barrat, Alain and Bonchi, Francesco and Cattuto, Ciro and Gullo, Francesco},
  title     = {Mining (Maximal) Span-cores from Temporal Networks},
  booktitle = {CIKM},
  year      = {2018}
}

@article{HeLYLJW24,
  author  = {He, Yue and Lin, Longlong and Yuan, Pingpeng and Li, Ronghua and Jia, Tao and Wang, Zeli},
  title   = {{CCSS}: Towards Conductance-Based Community Search with Size Constraints},
  journal = {Expert Systems with Applications},
  volume  = {250},
  pages   = {123915},
  year    = {2024}
}

@article{holme2012temporal,
  title   = {Temporal Networks},
  author  = {Holme, Petter and Saram{\"a}ki, Jari},
  journal = {Physics Reports},
  volume  = {519},
  number  = {3},
  pages   = {97--125},
  year    = {2012}
}

@inproceedings{huang2015minimum,
  title     = {Minimum Spanning Trees in Temporal Graphs},
  author    = {Huang, Silu and Fu, Ada Wai-Chee and Liu, Ruifeng},
  booktitle = {SIGMOD},
  pages     = {419--430},
  year      = {2015}
}

@INPROCEEDINGS{LiAWCS22,
  author    = {Li, Dengshi and Liang, Xiaocong and Hu, Ruimin and Zeng, Lu and Wang, Xiaochen},
  title     = {$(\alpha,\ \beta)$-AWCS: $(\alpha,\ \beta)$-Attributed Weighted Community Search on Bipartite Graphs},
  booktitle = {IJCNN},
  year      = {2022},
  pages     = {1--8}
}

@inproceedings{li2018persistent,
  title     = {Persistent Community Search in Temporal Networks},
  author    = {Li, Rong-Hua and Su, Jiao and Qin, Lu and Yu, Jeffrey Xu and Dai, Qiangqiang},
  booktitle = {ICDE},
  pages     = {797--808},
  year      = {2018}
}

@inproceedings{li2023persistent,
  title     = {Persistent Community Search Over Temporal Bipartite Graphs},
  author    = {Li, Mo and Xie, Zhiran and Ding, Linlin},
  booktitle = {ADMA},
  pages     = {324--339},
  year      = {2023}
}

@inproceedings{li2024historicalabc,
  author    = {Li, Shunyang and Wang, Kai and Lin, Xuemin and Zhang, Wenjie and He, Yizhang and Yuan, Long},
  title     = {Querying Historical Cohesive Subgraphs Over Temporal Bipartite Graphs},
  booktitle = {ICDE},
  pages     = {2503--2516},
  year      = {2024}
}

@ARTICLE{Linbil2025,
  author  = {Lin, Longlong and He, Yue and Chen, Wei and Yuan, Pingpeng and Li, Rong-Hua and Jia, Tao},
  title   = {Effective and Efficient Conductance-Based Community Search at Billion Scale},
  journal = {IEEE Transactions on Big Data},
  volume  = {11},
  number  = {6},
  pages   = {3170--3184},
  year    = {2025},
  doi     = {10.1109/TBDATA.2025.3588028}
}

@article{LinKDD24,
  author  = {Lin, Longlong and Jia, Tao and Wang, Zeli and Zhao, Jin and Li, Rong-Hua},
  title   = {{PSMC}: Provable and Scalable Algorithms for Motif Conductance Based Graph Clustering},
  journal = {KDD},
  pages   = {1793--1803},
  year    = {2024}
}

@article{LinQTCS24,
  author  = {Lin, Longlong and Yuan, Pingpeng and Li, Rong-Hua and Zhu, Chun-Xue and Qin, Hongchao and Jin, Hai and Jia, Tao},
  title   = {{QTCS}: Efficient Query-Centered Temporal Community Search},
  journal = {Proceedings of the VLDB Endowment},
  volume  = {17},
  number  = {6},
  pages   = {1187--1199},
  year    = {2024}
}

@article{LISize24,
  title   = {Maximal Size Constraint Community Search over Bipartite Graphs},
  journal = {Knowledge-Based Systems},
  volume  = {297},
  pages   = {111961},
  year    = {2024},
  author  = {Li, Mo and Borovica-Gajic, Renata and Choudhury, Farhana M. and Cui, Ningning and Ding, Linlin}
}

@article{liu2020efficient,
  title   = {Efficient ($\alpha$, $\beta$)-core Computation in Bipartite Graphs},
  author  = {Liu, Boge and Yuan, Long and Lin, Xuemin and Qin, Lu and Zhang, Wenjie and Zhou, Jingren},
  journal = {The VLDB Journal},
  volume  = {29},
  number  = {5},
  pages   = {1075--1099},
  year    = {2020}
}

@inproceedings{LiMoreliable,
  author    = {Li, Mo and Xie, Zhiran and Ding, Linlin},
  title     = {Reliable Community Search over Dynamic Bipartite Graphs},
  booktitle = {WISA},
  pages     = {298--307},
  year      = {2024}
}

@article{Luo2023coremain,
  author  = {Luo, Wensheng and Yang, Qiaoyuan and Fang, Yixiang and Zhou, Xu},
  title   = {Efficient Core Maintenance in Large Bipartite Graphs},
  journal = {Proceedings of the ACM on Management of Data},
  volume  = {1},
  number  = {3},
  pages   = {208:1--208:26},
  year    = {2023}
}

@ARTICLE{NCSAC2025,
  author={Lin, Longlong and Li, Quanao and Qiao, Miao and Wang, Zeli and Zhao, Jin and Li, Rong-Hua and Luo, Xin and Jia, Tao},
  journal={IEEE Transactions on Knowledge and Data Engineering}, 
  title={NCSAC: Effective Neural Community Search via Attribute-Augmented Conductance}, 
  year={2026},
  volume={38},
  number={2},
  pages={1221-1235},
 }

@inproceedings{qin2019mining,
  title     = {Mining Periodic Cliques in Temporal Networks},
  author    = {Qin, Hongchao and Li, Rong-Hua and Wang, Guoren and Qin, Lu and Cheng, Yurong and Yuan, Ye},
  booktitle = {ICDE},
  pages     = {1130--1141},
  year      = {2019}
}

@article{qin2020mining,
  title   = {Mining Stable Communities in Temporal Networks by Density-Based Clustering},
  author  = {Qin, Hongchao and Li, Rong-Hua and Wang, Guoren and Huang, Xin and Yuan, Ye and Yu, Jeffrey Xu},
  journal = {IEEE Transactions on Big Data},
  volume  = {8},
  number  = {3},
  pages   = {671--684},
  year    = {2020}
}

@article{qin2022mining,
  title   = {Mining Bursting Core in Large Temporal Graphs},
  author  = {Qin, Hongchao and Li, Rong-Hua and Yuan, Ye and Wang, Guoren and Qin, Lu and Zhang, Zhiwei},
  journal = {Proceedings of the VLDB Endowment},
  volume  = {15},
  number  = {13},
  pages   = {3911--3923},
  year    = {2022}
}

@article{SimaS06,
  author    = {Sima, Jiri and Schaeffer, Satu Elisa},
  title     = {On the NP-Completeness of Some Graph Cluster Measures},
  booktitle = {SOFSEM 2006: Theory and Practice of Computer Science},
  series    = {Lecture Notes in Computer Science},
  volume    = {3831},
  pages     = {530--537},
  publisher = {Springer},
  year      = {2006}
}

@inproceedings{shin2017densealert,
  author    = {Shin, Kijung and Hooi, Bryan and Kim, Jisu and Faloutsos, Christos},
  title     = {DenseAlert: Incremental Dense-Subtensor Detection in Tensor Streams},
  booktitle = {KDD},
  year      = {2017}
}

@article{wang2019vertex,
  title   = {Vertex Priority Based Butterfly Counting for Large-Scale Bipartite Networks},
  author  = {Wang, Kai and Lin, Xuemin and Qin, Lu and Zhang, Wenjie and Zhang, Ying},
  journal = {Proceedings of the VLDB Endowment},
  year    = {2019}
}

@article{wang2023significantbp,
  author  = {Wang, Kai and Zhang, Wenjie and Zhang, Ying and Qin, Lu and Zhang, Yuting},
  title   = {Discovering Significant Communities on Bipartite Graphs: An Index-Based Approach},
  journal = {IEEE Transactions on Knowledge and Data Engineering},
  volume  = {35},
  number  = {3},
  pages   = {2471--2485},
  year    = {2023}
}

@inproceedings{wang2024bipartite,
  title     = {Bipartite Graph Analytics: Current Techniques and Future Trends},
  author    = {Wang, Hanchen and Wang, Kai and Zhang, Wenjie and Zhang, Ying},
  booktitle = {ICDE},
  pages     = {01--07},
  year      = {2024}
}

@INPROCEEDINGS{wangICDE21,
  author    = {Wang, Kai and Zhang, Wenjie and Lin, Xuemin and Zhang, Ying and Qin, Lu and Zhang, Yuting},
  title     = {Efficient and Effective Community Search on Large-Scale Bipartite Graphs},
  booktitle = {ICDE},
  year      = {2021},
  pages     = {85--96}
}

@INPROCEEDINGS{wangperbc,
  author    = {Wang, Kai and Zhang, Wenjie and Lin, Xuemin and Qin, Lu and Zhou, Alexander},
  title     = {Efficient Personalized Maximum Biclique Search},
  booktitle = {ICDE},
  year      = {2022},
  pages     = {498--511}
}

@article{wu2014path,
  title   = {Path Problems in Temporal Graphs},
  author  = {Wu, Huanhuan and Cheng, James and Huang, Silu and Ke, Yiping and Lu, Yi and Xu, Yanyan},
  journal = {Proceedings of the VLDB Endowment},
  volume  = {7},
  number  = {9},
  pages   = {721--732},
  year    = {2014}
}

@article{wu2024frequency,
  author  = {Wu, Yanping and Sun, Renjie and Wang, Xiaoyang and Wen, Dong and Zhang, Ying and Qin, Lu and Lin, Xuemin},
  title   = {Efficient Maximal Frequent Group Enumeration in Temporal Bipartite Graphs},
  journal = {Proceedings of the VLDB Endowment},
  volume  = {17},
  number  = {11},
  pages   = {3243--3255},
  year    = {2024}
}

@article{Xvattrib,
  author  = {Xu, Zongyu and Zhang, Yihao and Yuan, Long and Qian, Yuwen and Chen, Zi and Zhou, Mingliang and Mao, Qin and Pan, Weibin},
  title   = {Effective Community Search on Large Attributed Bipartite Graphs},
  journal = {International Journal of Pattern Recognition and Artificial Intelligence},
  volume  = {37},
  number  = {2},
  pages   = {2359002:1--2359002:25},
  year    = {2023}
}

@INPROCEEDINGS{Zhangbitruss,
  author    = {Zhang, Zuxuan and Li, Hongxi and Huang, Kai and Jiang, Yuncheng},
  title     = {k-bitruss-Attributed Weighted Community Search on Attributed Weighted Bipartite Graphs},
  booktitle = {BigData},
  year      = {2023},
  pages     = {562--567}
}

@inproceedings{zhang2023discovering,
  title     = {Discovering Frequency Bursting Patterns in Temporal Graphs},
  author    = {Zhang, Qianzhen and Guo, Deke and Zhao, Xiang and Yuan, Long and Luo, Lailong},
  booktitle = {ICDE},
  year      = {2023}
}

@inproceedings{zhang2024size,
  author    = {Zhang, Yuting and Wang, Kai and Zhang, Wenjie and Ni, Wei and Lin, Xuemin},
  title     = {Size-bounded Community Search over Large Bipartite Graphs},
  booktitle = {EDBT},
  pages     = {320--331},
  publisher = {OpenProceedings.org},
  year      = {2024}
}

@article{zhou2021butterfly,
  title   = {Butterfly Counting on Uncertain Bipartite Graphs},
  author  = {Zhou, Alexander and Wang, Yue and Chen, Lei},
  journal = {Proceedings of the VLDB Endowment},
  volume  = {15},
  number  = {2},
  pages   = {211--223},
  year    = {2021}
}

@article{ZHOUSize23,
  title   = {Effective and Efficient Community Search with Size Constraint on Bipartite Graphs},
  journal = {Information Sciences},
  volume  = {647},
  pages   = {119511},
  year    = {2023},
  author  = {Zhou, Keqi and Xin, Junchang and Chen, Jinyi and Zhang, Xian and Wang, Beibei and Wang, Zhiqiong}
}

@article{LiCSS21,
  author  = {Li, Yuan and Liu, Jinsheng and Zhao, Huiqun and Sun, Jing and Zhao, Yuhai and Wang, Guoren},
  title   = {Efficient Continual Cohesive Subgraph Search in Large Temporal Graphs},
  journal = {World Wide Web},
  volume  = {24},
  number  = {5},
  pages   = {1483--1509},
  year    = {2021}
}

@article{LIactive,
  author  = {Li, Ling and Zhao, Yuhai and Li, Yuan and Wahab, Fazal and Wang, Zhengkui},
  title   = {The Most Active Community Search in Large Temporal Graphs},
  journal = {Knowledge-Based Systems},
  volume  = {250},
  pages   = {109101},
  year    = {2022}
}

@inproceedings{ZhangEngage22,
  author    = {Zhang, Yifei and Lin, Longlong and Yuan, Pingpeng and Jin, Hai},
  title     = {Significant Engagement Community Search on Temporal Networks},
  booktitle = {DASFAA},
  pages     = {250--258},
  year      = {2022}
}

@article{Yangtruss,
  author  = {Yang, Huihui and Zhu, Chunxue and Lin, Longlong and Yuan, Pingpeng},
  title   = {Towards Truss-Based Temporal Community Search},
  journal = {CoRR},
  volume  = {abs/2410.15046},
  year    = {2024}
}

@article{DuMTZ23,
  author  = {Du, Ming and Ma, Wanting and Tan, Yuting and Zhou, Junfeng},
  title   = {Continuous Community Search with Attribute Constraints in Temporal Graphs},
  journal = {The Journal of Supercomputing},
  volume  = {79},
  pages   = {15734--15759},
  year    = {2023}
}

@article{Zhong25kcore,
  author       = {Zhi Wang and
                  Ming Zhong and
                  Yuanyuan Zhu and
                  Tieyun Qian and
                  Mengchi Liu and
                  Jeffrey Xu Yu},
  title        = {On More Efficiently and Versatilely Querying Historical k-Cores},
  journal      = {Proc. {VLDB} Endow.},
  volume       = {18},
  number       = {5},
  pages        = {1335--1347},
  year         = {2025},
}

@inproceedings{spi2004,
author = {Spielman, Daniel A. and Teng, Shang-Hua},
title = {Nearly-linear time algorithms for graph partitioning, graph sparsification, and solving linear systems},
booktitle = {STOC},
}

@article{tabc24,
  author       = {Anxin Tian and
                  Alexander Zhou and
                  Yue Wang and
                  Xun Jian and
                  Lei Chen},
  title        = {Efficient Index for Temporal Core Queries over Bipartite Graphs},
  journal      = {Proc. {VLDB} Endow.},
  volume       = {17},
  number       = {11},
  pages        = {2813--2825},
  year         = {2024},
}

@article{zhang2150top,
  title={Top-r Influential Community Search in Bipartite Graphs},
  author={Zhang, Yanxin and Hua, Zhengyu and Yuan, Long and Chen, Zi},
  journal={Proceedings of the VLDB Endowment},
  volume={2150},
  pages={8097},
  year={2025}
}

@article{Beyerrobust07,
title = {Robust optimization – A comprehensive survey},
journal = {Computer Methods in Applied Mechanics and Engineering},
volume = {196},
number = {33},
pages = {3190-3218},
year = {2007},
issn = {0045-7825},
author = {Hans-Georg Beyer and Bernhard Sendhoff}
}

@article{Gor15,
title = {A practical guide to robust optimization},
journal = {Omega},
volume = {53},
pages = {124-137},
year = {2015},
issn = {0305-0483},
author = {Bram L. Gorissen and İhsan Yanıkoğlu and Dick {den Hertog}}
}
